# E-WAN: Efficient Communication in Energy Harvesting Low-Power Networks

NAOMI STRICKER, ETH Zurich, Switzerland
DAVID BLASER, ETH Zurich, Switzerland
ANDRES GOMEZ, TU Braunschweig, Germany
LOTHAR THIELE, ETH Zurich, Switzerland

The ever-increasing number of distributed embedded systems in the context of the Internet of Things (IoT), Wireless Sensor Networks (WSN), and Cyber-Physical Systems (CPS) rely on wireless communication to collect and exchange data. Nodes can employ single-hop communication which, despite its ease, may necessitate energy-intensive long-range communication to cover long distances. Conversely, multi-hop communication allows for more energy-efficient short-range communication since nodes can rely on other nodes to forward their data. Yet, this approach requires relay nodes to be available and continuous maintenance of a dynamically changing distributed state. At the same time, energy harvesting has the potential to outperform traditional battery-based systems by improving their lifetime, scalability with lower maintenance costs, and environmental impact. However, the limited and temporally and spatially variable harvested energy poses significant challenges for networking in energy harvesting networks, particularly considering the energy demands and characteristics of both multi-hop and single-hop communication. We propose E-WAN, a protocol for energy harvesting wide-area low-power networks that builds on the concept of *virtual sub-networks* to enable resource-efficient multi-hop communication when possible and reliable however energy-intensive point-to-point communication otherwise. Nodes autonomously and dynamically move between the two and adjust to changing network states and resources based only on easily obtainable network state information. We illustrate E-WAN's advantages both in terms of efficiency and adaptability in various communication and harvesting scenarios. Furthermore, we demonstrate E-WAN operating in a realistic setting by deploying an energy harvesting network in a real-world indoor environment.

## 1 INTRODUCTION

With the rise of the Internet of Things (IoT), Wireless Sensor Networks (WSN), and Cyber-Physical Systems (CPS), distributed embedded systems have become increasingly pervasive, finding applications in diverse fields including for example building control [35] or environmental monitoring [67]. These network typically consists of distributed nodes with sensing, processing and communication capabilities and a gateway. Since powering the nodes with mains-power is typically not feasible also not in indoor applications due to various factors such as poor scalability and costs associated with cables, only the gateway is mains-powered. Consequently, the distributed nodes rely on alternative power sources. Recent research has explored powering nodes with photovoltaic energy harvesting as a promising alternative to battery-operated devices, even in energy-scarcer indoor environments. Energy harvesting-based networks offer extended lifetimes, improved scalability with lower maintenance costs, and a lower environmental impact compared to traditional battery-operated networks.

Efficient wireless communication is essential for networks to enable resource-constrained nodes to collect, distribute, and share data. To support applications extending over wide areas, nodes can use short-range communication and rely on other resource-constrained nodes to forward their data, thereby bridging larger distances. These multi-hop networks typically require distributed network state information, which, depending on the protocol, may include routing trees, time synchronization, link budgets, neighbor nodes, or network topology [3, 71]. However, maintaining this distributed state is challenging, as it changes dynamically e.g., due to mobile nodes or fluctuations in the network's environment and thus RF propagation. Therefore, the state needs to be continuously updated, and networking protocols need to adapt accordingly. Meanwhile, advances in low-power wide-area networks (LPWANs) have extended the connectivity of nodes, enabling low-power long-range communication technologies that facilitate simple star-based topologies, e.g., LoRa [43]. Single-hop communication with the gateway allows nodes to communicate independently of the network topology. Yet, despite the advances, these protocols generally incur a substantially higher energy cost per communicated packet compared to multi-hop communication [45, 62].

While communication in resource-constrained networks is inherently difficult, energy harvesting networks encounter distinct and significant challenges, particularly in real-world indoor environments. In such settings, energy harvesting provides only small amounts of energy characterized by unpredictable temporal and spatial variability [31]. As a result, low-power networking protocols for energy harvesting networks face major difficulties, both due to the limited, variable, and volatile energy that nodes harvest and because of the energy requirements, properties, and variability of both multi-hop and single-hop communication.

Harvesting-based nodes have unpredictable and limited resources available, which may cause them to run out of energy frequently. Therefore, they may be unable to communicate continuously or even maintain the necessary state information for multi-hop communication [58]. Furthermore, when and for how long each node can communicate is not known in advance and can differ substantially between nodes. Specifically, energy harvesting nodes frequently but unpredictably run out of power, temporarily ceasing their active participation in network formation and communication. As they regain power at unpredictable times, they can rejoin and actively participate and communicate in the network. This leads to a network with dynamically changing size and topology consisting only of the gateway and nodes that are powered and communicating at a given time. An illustrative example is depicted in Figure 1. Consequently, communication needs to support nodes with no information of neighboring nodes' existence, state, or the timing of the network communication to efficiently bootstrap, i.e., join the network. In addition, such scenarios require *dynamic adaptability*: The wireless communication protocol must enable nodes to dynamically adapt to changing network size, state, and topology steming from fluctuating link characteristics and variations in node availabilities. At the same time resource *efficiency* in terms of successfully transmitted packets relative to harvested energy is crucial. Efficiency is influenced by various factors, including the resource requirements of communication, overheads incurred by bootstrapping, reliability, and a node's ability to communicate in different network states.

Efficiency and dynamic adaptability are also key in addressing communication challenges in battery-based networks. Yet, major differences in the characteristics of their resources inhibit directly applying approaches designed for battery-based networks to energy harvesting networks. Battery-based nodes have a stable power source, allowing them to maintain network information and flexibly adjust energy usage throughout their lifetime. Even when node mobility or weak links cause nodes to lose connectivity, they are still able to operate, to maintain network information, and may even be able to overcome link losses with longer-range communication. Once their battery is depleted, they become and remain unreachable. Protocols for such networks only take into account operating and reachable nodes. In contrast, energy harvesting nodes are restricted in their ability to temporally shift energy usage due to their limited energy storage capacity, thus requiring efficient and effective use of available resources and swift adaptations to network changes. Furthermore, unlike battery-based nodes, energy harvesting nodes may frequently run out of energy and therefore cannot continuously maintain network information, operate or remain reachable. They disconnect from the network and are unable to address this disconnection until they accumulate sufficient energy and become relevant for networking again.

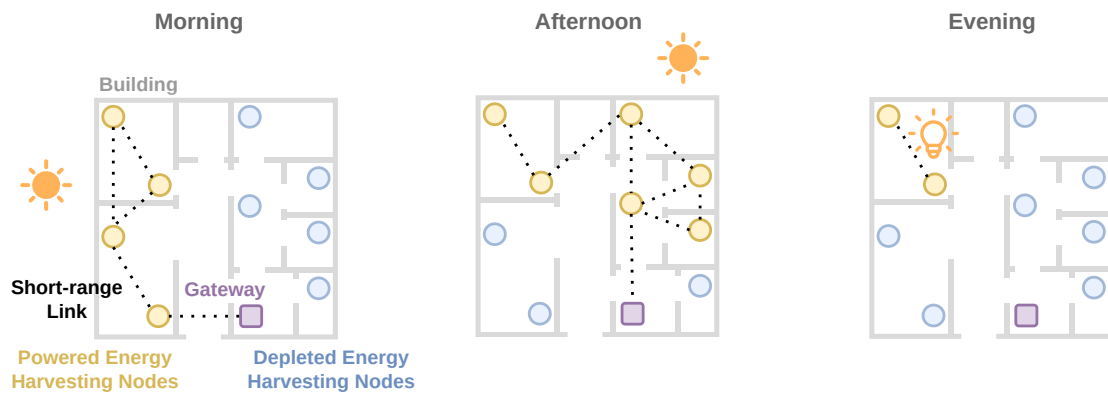

Fig. 1. Variability of photovoltaic energy harvesting in energy-scarce indoor environments impacts which nodes have energy and when and thus the size and topology of the network formed by currently powered nodes. Long-range communication can provide reliable point-to-point links from any node to the gateway. It is, however, less efficient than short-range multi-hop communication which in turn is affected by the dynamically changing network state.

Nonetheless, various approaches have been proposed to address similar challenges in both battery-based and energy harvesting-based wireless networks. A detailed review is given in Section 7. Changes in network topology impact connectivity in both networking scenarios, for example, affecting highly mobile battery-based network devices such as UAVs communicating either within a swarm [21] or with a ground station [68], as well as static energy harvesting-based network devices, whose dynamic energy availability dictates available communication links. Point-to-point communication with the gateway enables energy harvesting nodes to communicate based on local resources, independent of network topology. This approach has been employed with efficient short-range communication [60], which covers only small areas, and with energy-intensive long-range communication for wide-area networks [22]. Larger distances can also be bridged with multi-hop communication, for which time synchronization is important [55]. Various approaches have been proposed to enable harvesting-based nodes to efficiently bootstrap after power losses, e.g., [30, 58]. In [58], we propose a bootstrapping method that leverages asynchronous long-range point-to-point communication and facilitates energy harvesting nodes efficiently joining a short-range multi-hop network. Existing multi-hop communication protocols for energy harvesting networks typically optimize communication based on energy predictions and link characteristics [2]. Yet, indoor energy harvesting is difficult to predict, and obtaining sufficiently accurate link characteristics for the dynamically changing network implies a significant overhead. These approaches, furthermore, do not address cases when powered nodes do not have a multi-hop path to the gateway. Similar scenarios in battery-based networks due to inhomogeneous links have been addressed in recent work by combining short- and long-range communication [64].

This work builds on the bootstrapping approach we proposed in [58] and presents the following major contributions to overcoming the networking challenges of energy harvesting networks in real-world environments:

- We design a communication protocol, E-WAN, for energy harvesting wide-area low-power networks that leverages both short- and long-range communication. It builds on the concept of virtual sub-networks and enables nodes to dynamically and autonomously adjust to changing network states and their energy availability. Furthermore, we provide properties of the protocol in terms of its efficiency and dynamic adaptability.

- We extensively evaluate the proposed protocol in comparison to a single-hope baseline that relies solely on long-range point-to-point communication and a multi-hop baseline that employs efficient short-range multi-hop communication. The protocols are evaluated for two energy harvesting scenarios and four network topologies using discrete event simulations. Furthermore, we implement the proposed protocol on real hardware. We conduct a case study in an indoor environment, where we deploy and measure a small energy harvesting network to demonstrate the protocol operating under realistic conditions and validate the discrete event simulations.

The remainder of this work is structured as follows: Section 2 provides an overview of energy harvesting nodes, LPWANs, and the considered scenarios. The proposed protocol and its properties are presented in Section 3 and Section 4, respectively. Section 5 summarizes the implementation of E-WAN, and Section 6 presents an extensive evaluation using simulations and a real-world case study. Subsequently, we review related work in Section 7 and conclude in Section 8.

## 2 PRELIMINARIES

In this section, we provide an overview of energy harvesting nodes and wireless communication in LPWANs. Furthermore, we describe the general scenario for which the proposed protocol is designed.

### 2.1 Energy Harvesting Nodes

Understanding the characteristics and operation of energy harvesting nodes is crucial for identifying and addressing communication challenges inherent to energy harvesting networks. Harvesting-based nodes typically follow a harvest-store-use architecture, which is illustrated in Figure 2a. Nodes rely on a transducer to convert primary energy such as radiation or thermal energy into electrical energy. Among transducers, photovoltaic cells offer a comparably high power density, even in typically energy-scarce indoor environments [2, 10]. The harvested energy is stored in an energy storage element such as a supercapacitor. The application, which typically performs sensing, processing, and wireless communication tasks, is powered by the stored energy. A node's energy autonomy is tied to the capacity of its energy storage. Batteryless energy harvesting nodes, e.g., [54], have minimized energy storages, while nodes with larger storage capacities can sustain themselves for a limited time, bridging several hours or days without energy input, e.g., [48]. Nodes with a larger energy storage rely on energy management strategies to control resource usage, adapting their operation, such as their duty cycle, in accordance with their available resources. This dynamic behavior is based on parameters such as the state of charge and estimations of future harvested energy [2]. Although first advances in predicting short-term variability of photovoltaic indoor energy harvesting have been presented, e.g., in [59], forecasting energy availability over longer time horizons remains a significant challenge due to the inherent non-deterministic temporal and spatial variability of indoor photovoltaic energy harvesting. Figure 2b shows two example harvesting traces for a solar panel with dimensions of 50 mm × 33 mm, recorded over the same time frame. The traces are part of a dataset collected over multiple years in an office building [53]. Despite being measured merely a few meters apart, the traces exhibit significantly different characteristics, providing different amounts of energy at different times to the nodes.

### 2.2 Low-Power Wide-Area Networks

Low-power wide-area networks (LPWANs) are an efficient and effective approach to connecting large numbers of resource-constrained nodes distributed throughout a wide area. Wireless communication in LPWANs can cover long distances by employing long-range communication with which

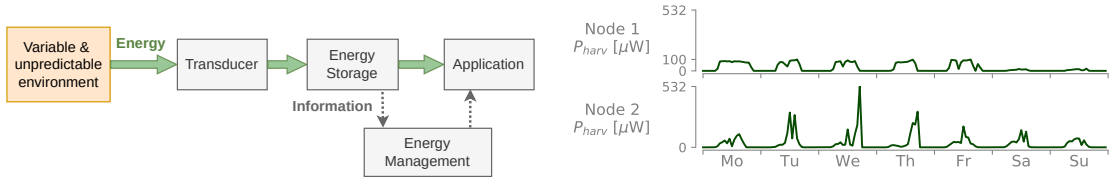

(a) In a harvest-store-use architecture, energy harvested by a transducer is stored in an energy storage which provides the energy to the application.

(b) Nodes at different locations in an office building can have considerably different harvesting characteristics.

Fig. 2. Photovoltaic energy harvesting nodes commonly have a harvest-store-use architecture. With the energy storage element, nodes can bridge limited time periods with no input energy. They adapt their behavior to their local resource availability which in indoor environments varies unpredictably.

nodes have extended connectivity. Several low-power long-range modulation schemes, typically with low bandwidth, have been developed, such as LoRa [43]. With this improved sensitivity, nodes in an LPWAN can have direct links to the gateway, forming a star topology. Various protocols for such single-hop networks have been developed, including LoRaWAN [1]. While star topologies facilitate asynchronous communication, time-synchronized approaches have also been proposed to avoid packet collisions.

Large communication ranges can also be achieved with more efficient short-range communication whereby nodes depend on others to forward their data. Protocols for multi-hop networks can exploit information about the network state, communication demand, topology, or link budgets to schedule communication and route packets through the network [3]. A specific class of protocols relies on synchronous transmissions, where nodes purposely retransmit received data simultaneously. For example, in Glossy [28] and Gloria [64], a node sends its packet while all other nodes listen and, upon reception, retransmit the packet at synchronized times, thus flooding the network with the data. These flooding-based protocols are typically characterized by low latency, high reliability, and robustness because they do not rely on the state of the network topology [71].

### 2.3 Scenario

The proposed protocol is designed for networks where $N$ energy harvesting nodes and a single mains-powered gateway, hereafter referred to as host, form an LPWAN. Both the energy harvesting nodes and the host are static. The host is assumed to have no relevant resource constraints and may additionally be connected to the Internet. The energy harvesting nodes harvest small amounts of energy from temporally and spatially varying primary energy in their environments and include a sufficiently large energy storage, enabling them to bridge several hours without harvesting any energy. All nodes require only a low communication bandwidth to exchange relevant information, such as sensor measurements. These communication demands may differ between the nodes in the network.

The host and the harvesting-based nodes support multiple modulation schemes that enable communication over short and long distances. An overview of the scenario is shown in Figure 3. Short-range communication is energy efficient, yet not all nodes have a direct short-range link to the host and instead rely on other energy harvesting nodes in the network to forward their data. There are no dedicated relay nodes with more resources, only energy harvesting nodes are available to retransmit packets from other energy harvesting nodes. With long-range communication, a direct link between each energy harvesting node and the host is possible.

Due to fluctuating link characteristics and unpredictable and variable energy availability, nodes regularly run out of energy and stop actively participating in the network's communication until they have harvested enough energy again. This results in a dynamically changing network size, state, and topology, as a node may or may not be connected to the currently existing multi-hop network and may or may not have sufficient energy to use the single-hop long-range connection to the host. Wireless communication in energy-harvesting LPWANs must be resilient and adapt to such changes. Consequently, a communication protocol for energy harvesting networks must enable nodes to *dynamically adapt* to network changes. In particular, energy harvesting nodes need to dynamically choose between long-range and short-range communication based on varying networking conditions. Furthermore, due to the limited resources of energy harvesting nodes, it is important for communication, as well as any overhead associated with dynamic behavior, to be *efficient*. Such efficiency is required to enable nodes to communicate a large number of packets with the limited and variable energy they harvest. This resource efficiency has to be considered for each energy harvesting node individually, since resources are not shared and nodes have energy storages with limited capacity.

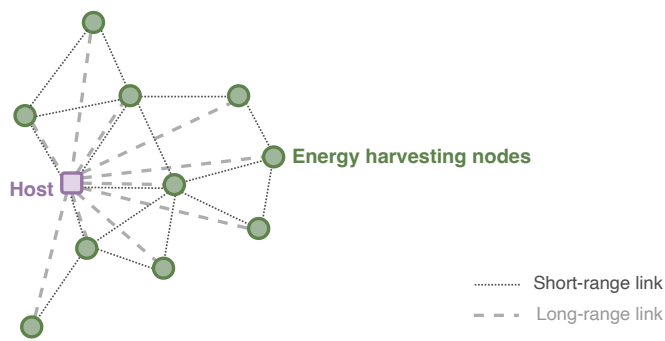

Fig. 3. The proposed protocol is designed for distributed embedded systems comprised of $N$ energy harvesting nodes and a mains-powered host. Every node can communicate directly with the host with long-range communication which is energy-intensive. Multi-hop communication that utilizes short-range links is more energy-efficient but requires that a multi-hop path from the host to a node with communication demand is connected. However, due to various challenges associated with energy harvesting networks, this might not always be the case.

## 3 E-WAN: COMMUNICATION PROTOCOL FOR ENERGY HARVESTING LOW-POWER NETWORKS

In this section, we present E-WAN, the proposed communication protocol for energy harvesting networks. It enables nodes in the network to efficiently communicate and dynamically adapt to varying resources and network topologies while imposing only small overheads. Additionally, nodes maintain only little information about the network state. We first provide an overview of the general concept of E-WAN and subsequently describe it in detail. The properties of E-WAN are elaborated on in Section 4.

### 3.1 General Concept of E-WAN

E-WAN builds on the concept of virtual sub-networks (VSNs), where physical devices in a network are logically organized in groups referred to as virtual sub-networks. Devices in a VSN communicate according to the sub-network's communication protocol, specifications, and parameters. Clustering in networks commonly refers to an approach where nodes are also grouped, typically with a selected cluster head that coordinates communication within the cluster and relays data to the gateway through other cluster heads, introducing hierarchy in the communication [70]. In contrast, each VSN in the proposed protocol functions as a separate, centrally coordinated network with the gateway, enabling flexible and scalable network designs.

E-WAN incorporates VSNs to address the challenges of energy harvesting low-power wireless networks. It consists of three VSNs that operate concurrently, each designed to address specific communication challenges in energy harvesting networks. The number of VSNs is selected to achieve a balance between addressing communication challenges and the resulting complexity of the protocol.

The host is always part of all three VSNs and energy harvesting nodes belong exclusively to one VSN while they have enough energy to communicate. Harvesting-based nodes can logically leave, join, and move between the VSNs according to their resources and changes in the network state. Two VSNs, a single-hop VSN and a multi-hop VSN, are designed for data collection and dissemination. The former provides reliable but energy-intensive data communication, while the latter is energy-efficient but not always possible. The third VSN, a bootstrapping VSN, enables nodes to efficiently join the single-hop and multi-hop VSNs.

Communication in the single-hop and multi-hop VSNs occurs in periodic rounds and follows a time-division multiple-access (TDMA) approach. The host determines a separate global schedule for each and schedules the rounds for the two VSNs in sequence. This temporally isolates the rounds and ensures that they do not interfere with each other. The multi-hop VSN employs energy-efficient short-range communication, which is essential for resource-constrained networks. Yet, nodes may depend on other energy harvesting nodes to relay their data, and these nodes may not be part of this VSN. This can arise, for example, due to the limited resources of harvesting-based nodes, which cause nodes to run out of energy regularly. We observed in [58] that in indoor environments, energy harvesting nodes frequently experience power losses even when they have relatively large system dimensions and only rely on energy-efficient short-range multi-hop communication. The single-hop VSN relies on point-to-point communication between each energy harvesting node and the host to provide reliable and guaranteed but energy-intensive communication. It is designed to overcome topological sparsity and disconnections in the network's short-range topology.

The bootstrapping VSN is based on the approach we proposed in [58] and facilitates nodes joining one of the other two VSNs with extremely limited and predictable energy cost, as demonstrated in [58]. Through an asynchronous point-to-point packet exchange with the host, nodes have direct and immediate access to timing and coordination information for communication in the single-hop and multi-hop VSNs. With this information, nodes turn off their radios to save energy until the next relevant network activity, at which point they wake up and efficiently scan for network traffic.

Energy harvesting nodes can leave, join, or move from one VSN to another depending on their local energy resources and varying network states. For example, nodes transition from the multi-hop to the single-hop VSN if they no longer receive multi-hop network traffic. Conversely, nodes in the single-hop VSN periodically scan for network traffic from the multi-hop VSN and move to it if they successfully receive the multi-hop VSN's global schedule, which indicates the existence of a short-range multi-hop path from the host to the node in the multi-hop VSN.

An overview of E-WAN's VSNs with their resource requirements and topological dependencies and the transitions between the VSNs are depicted in Figure 4. The communication in each VSN, their concurrent operation, and the transitions between VSNs are described in more detail in the following.

### 3.2 Communication in VSNs

The communication in each VSN is designed and optimized for its intended functionality and characteristics.

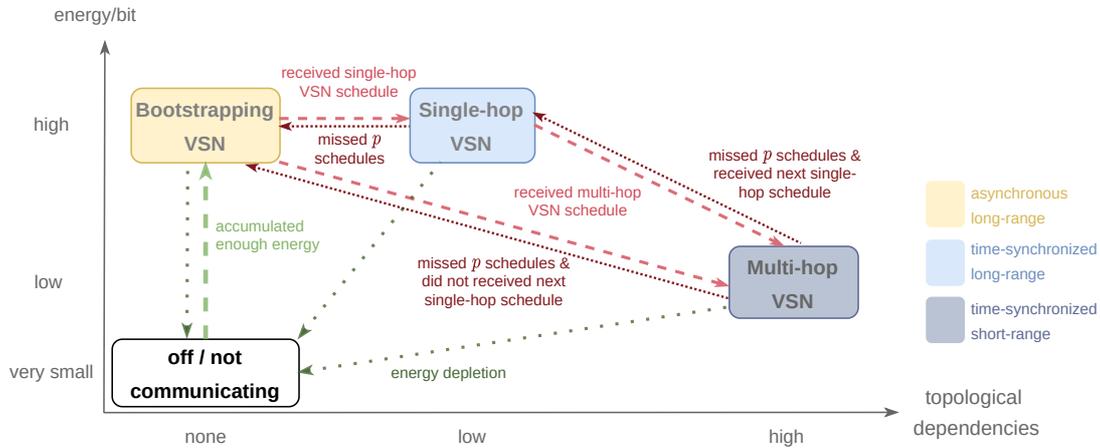

Fig. 4. The communication resource requirements and the topological dependencies energy harvesting nodes have on other nodes for the three VSNs and the transitions between VSNs. Nodes that are not communicating join the bootstrapping VSN once they accumulate enough energy to communicate and nodes that run out of energy stop communicating. Between the VSNs, nodes transition based on receiving and missing schedules. They are thus able to autonomously adapt to changes in a VSN's network state and locally optimize their energy requirements to communicate data.

*Multi-hop VSN.* The multi-hop VSN is optimized for energy-efficient data collection and dissemination. It relies on multi-hop communication allowing it to employ efficient short-range communication and cover wide areas by relying on nodes to forward data from other nodes. Energy harvesting nodes optimize the use of their resources by joining the multi-hop VSN whenever possible. A node is able to join the multi-hop VSN if a multi-hop path from the host to the node exists in the multi-hop VSN, i.e. when it is connected to the VSN's multi-hop sub-network. Consequently, for communication in the multi-hop VSN, nodes need to form a connected graph. When the communication parameters in the multi-hop VSN result in relatively large link budgets, this can typically be fulfilled with a sparser network than if the link budgets are smaller. Nodes remain in this VSN as long as they receive schedules from the host and only leave if they miss $p$ consecutive schedules. A detailed description of the parameter $p$ and the dynamic transitions between VSNs is given in Section 3.3.

Communication in the multi-hop VSN takes place in periodic rounds with a constant period $T$ according to a global schedule determined by the host. A round in the multi-hop VSN follows the structure and design of a communication round in the Low-power Wireless Bus (LWB) protocol [27]. An overview of a round is depicted in Figure 5. Rounds consist of non-overlapping slots in which at most one node sends a message. Packets are transmitted with Gloria floods [64], where a node sends a message by initiating a flood, and all other nodes in the multi-hop VSN listen and, upon reception, retransmit the packet at the same time. By relying on flooding in the multi-hop VSN, its communication is robust and resilient, facilitating nodes to communicate in this VSN whenever possible. In the first slot of a round, the host sends the schedule of the current round. The schedule also contains the timing and networking information of the single-hop VSN needed for nodes to join the single-hop VSN. The host can include this information since it coordinates communication in both VSNs. This enables nodes to transition between VSNs as described in detail in Section 3.3. The schedule slot is followed by a variable number of data slots assigned to nodes.

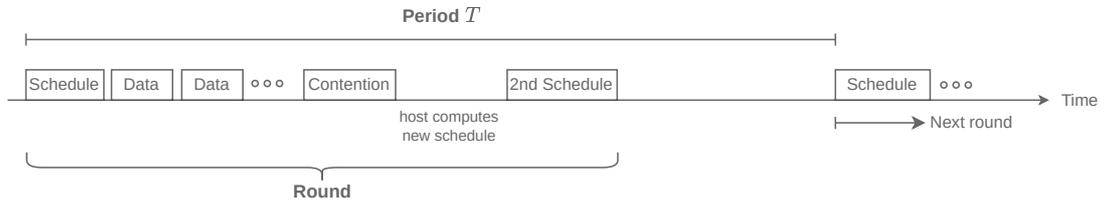

Fig. 5. Communication in the multi-hop and single-hop VSN occurs in rounds with a period $T$. The host begins a round by sending the round's schedule. Subsequently, nodes send their data and in the contention slot, they can inform the host of changes in their traffic demands. After updating the schedule, the host sends a second schedule.

The maximum number of data slots that can be included in a round is limited and depends, for example, on the protocol's timing, as elaborated in the section on the concurrent operation of VSNs. Subsequently, during a contention slot, any node in the VSN can inform the host of its traffic demands. The host computes a new schedule incorporating the previously received communication demand and removing any data slots assigned to nodes from which it has not received any data for $p$ consecutive rounds, see Section 3.3. The final slot is reserved for a second schedule sent by the host. The second schedule encompasses timing information for the upcoming round and changes to the schedule. It enables nodes to efficiently plan their operation ahead of time, turn off their transceiver, and enter an energy-efficient state accordingly.

Since the multi-hop VSN uses short-range communication, its communication is energy-efficient. Furthermore, no knowledge of the network state is required as in LWB. However, nodes can only communicate in this VSN if they are connected to its currently existing multi-hop network.

*Single-hop VSN.* The single-hop VSN is designed for reliable and guaranteed, albeit energy-intensive, communication. It relies on point-to-point communication between any harvesting-based node and the host, which typically requires long-range modulation schemes such as LoRa. This results in the nodes and the host forming a network with a star topology, where any node has a single-hop connection to the host and at most a two-hop connection to any other node via the host. The simple topology comes at a higher resource cost for sending a packet compared to the multi-hop VSN. Energy harvesting nodes remain in this VSN until they either move to the multi-hop VSN or miss $p$ consecutive schedules, see Section 3.3.

Communication follows a TDMA approach, where slots for nodes are grouped into periodic rounds with a constant period of $T$, equal to the period of communication in the multi-hop VSN. A round has the same structure as a multi-hop round, depicted in Figure 5. The schedule sent by the host includes timing and networking information of the multi-hop VSN communication, enabling nodes to transition to the multi-hop VSN, see Section 3.3. The single-hop topology simplifies packet transmission and reception in slots. All nodes in this VSN directly receive the schedules sent by the host. During data slots, a node sends its packet and the host repeats it to disseminate the message. Equivalent to the multi-hop VSN, the maximum number of data slots that can be allocated in the single-hop VSN is limited.

*Bootstrapping VSN.* The primary function of the bootstrapping VSN is to enable nodes to efficiently join either the multi-hop or single-hop VSN. Bootstrapping is necessary for nodes to synchronize and know when relevant network traffic occurs. For example, when nodes run out of energy, they lose time synchronization and must bootstrap once they have harvested enough energy to begin communicating again.

Communication in this VSN follows the efficient bootstrapping mechanism we proposed in [58]. An overview of the bootstrapping VSN is depicted in Figure 6. A node in the bootstrapping VSN sends an asynchronous point-to-point synchronization request to the host. The packet indicates to the host that a node is bootstrapping and needs timing and coordination information for the other two VSNs. Between communication rounds of the single-hop and multi-hop VSN, the host idle listens for synchronization requests. Upon reception of a valid request packet, the host responds with the necessary information for the node to join both the multi-hop and single-hop VSNs. The host sends the time until the next multi-hop communication $\tau_1$ and until the subsequent single-hop communication $\tau_2$, i.e., the single-hop round that occurs after the multi-hop communication such that $\tau_1 < \tau_2$. With this information, the node enters a deep sleep state with its transceiver turned off until the next relevant multi-hop VSN traffic occurs, i.e., for a duration of $\tau_1$. Subsequently, the node scans for network traffic from the multi-hop VSN. If the node receives the schedule of the multi-hop communication, it joins the multi-hop VSN and immediately participates in the multi-hop round, where it can send its communication demand in the contention slot. If the node does not receive a valid schedule during the multi-hop schedule slot, it returns to an energy-efficient state until the known occurrence of the subsequent single-hop round. The bootstrapping node then listens for the single-hop VSN's schedule slot. Upon receiving the schedule, the node joins the single-hop VSN and announces its traffic demands in the round's contention slot. If a node fails to receive the schedule of both the multi-hop VSN and the single-hop VSN, or does not receive any networking information from the host, it attempts to bootstrap again later by sending another synchronization request packet. The timing of the next bootstrapping attempt depends on a node's energy availability and a randomized back-off time. A node may not receive the packet from the host for various reasons, for example, the host could be unavailable because it is participating in multi-hop or single-hop rounds or due to simultaneous synchronization requests. There is a trade-off between the time dedicated to rounds and when the host is available for asynchronous exchanges.

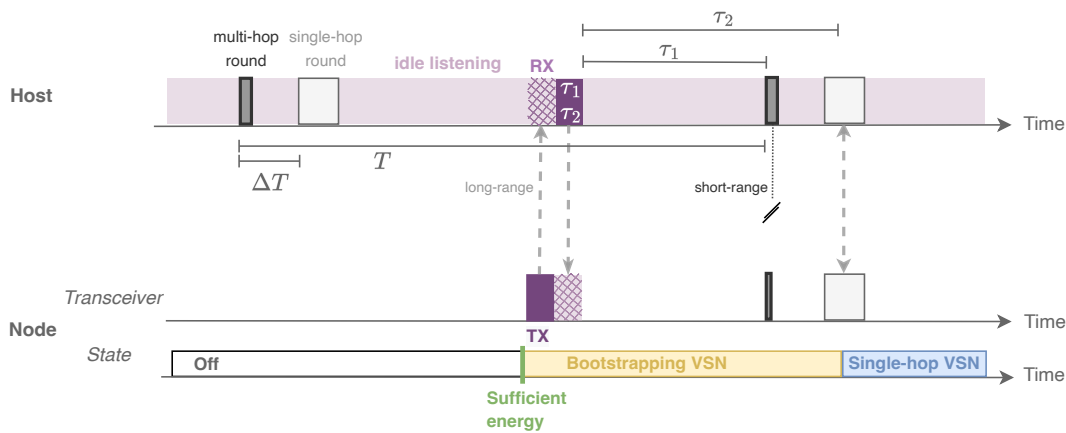

Fig. 6. Single-hop rounds are scheduled $\Delta T$ after multi-hop rounds both of which have the same period $T$. Between the rounds, the host idle listens for asynchronous long-range bootstrapping requests. When a node has sufficient energy to start communicating, it joins the bootstrapping VSN and sends an asynchronous long-range request. The host sends timing and networking information of the next multi-hop round, at time $\tau_1$ after the request and the subsequent single-hop round at $\tau_2$ where $\tau_1 < \tau_2$. The node enters a deep sleep state for time $\tau_1$ and then listens for a schedule. If the node is unable to receive a packet in the multi-hop schedule slot, it waits in deep sleep until the single-hop schedule slot. Upon successfully receiving a schedule, it joins the respective VSN.

The asynchronous point-to-point communication with the host gives bootstrapping nodes direct access to networking information and enables them to spend most of the bootstrapping time in an energy-efficient state. Consequently, the energy nodes require to bootstrap depends linearly on $\tau_1$ and $\tau_2$ increasing only at the minuscule rate of a node's constant power dissipation in its sleep state [58]. Furthermore, by first scanning the multi-hop VSN, nodes rely on energy-efficient communication when possible, see also Section 4.

*Concurrent Operation of VSNs.* An overview of how the three VSNs operate concurrently is depicted in Figure 6. Since the host coordinates the periodic rounds in the single-hop and multi-hop VSN, it can temporally separate them by scheduling one after the other. It schedules the rounds in the single-hop VSN to start $\Delta T$ after the beginning of the rounds in the multi-hop VSN. $\Delta T$ is a constant, that is sufficiently large to prevent any overlap between the two rounds. This ensures that the two do not interfere with each other and do not impact each other's performance. The period $T$ and the time difference $\Delta T$ between multi-hop and single-hop rounds need to be dimensioned in conjunction with the maximum number of data slots each round in the respective VSN can support. Considering the duration of a data slot, which depends on the communication parameters of the respective VSN, the timing of the rounds needs to be designed to ensure they are non-overlapping. Between rounds, the host idle listens for synchronization requests in the bootstrapping VSN. The bootstrapping VSN leverages asynchronous long-range communication initiated by the energy harvesting node, which is essential for enabling nodes to efficiently join and start communicating in the other two VSNs. However, bootstrapping nodes are not synchronized to the network-wide global time and therefore may send requests during the rounds of the other two VSNs. To isolate the communication in the bootstrapping VSN, particularly requests sent asynchronously by nodes from the communication rounds, different channels are used in all three VSNs.

## 3.3 Transitions between VSNs

While the host is always part of all three VSNs, which is feasible since we assume it is not harvesting-based, energy harvesting nodes are in one of the states depicted in Figure 4 and thus belong to at most one VSN. Furthermore, energy harvesting nodes can leave, join, and move between VSNs to dynamically adapt to their resources and changing network states. An overview of the transitions is shown in Figure 4 and an example of nodes in a network dynamically adapting to changing network states and resource availabilities is illustrated in Figure 7.

*Energy Driven Transitions.* Transitions due to a node's resource availability follow its energy management strategy. When a node runs out of energy and turns off or when its energy management determines it should stop communicating, it ceases to communicate and leaves whichever VSN it is currently in. Once the energy management determines that the node has accumulated sufficient stored energy to start communicating again, the node joins the bootstrapping VSN.

*Bootstrapping to Multi-hop or Single-hop VSN.* If a bootstrapping node receives networking information from the host, it can efficiently scan for network traffic from the multi-hop and single-hop VSNs. A node moves from the bootstrapping VSN to the multi-hop VSN if it receives the schedule of the multi-hop VSN. A newly joined node in the multi-hop VSN immediately participates in the flood of the schedule to accelerate the joining of other nodes to the multi-hop VSN. If a node does not receive the multi-hop schedule, it scans for the subsequent single-hop schedule and, upon receiving it, joins the single-hop VSN.

*Multi-hop to Single-hop or Bootstrapping VSN.* Transitions between the multi-hop and single-hop VSNs depend on the network state. Topology-induced transitions can result from changing RF environments or other energy harvesting nodes leaving and joining VSNs. Nodes remain in the multi-hop VSN until they miss $p$ consecutive schedules. After $p$ consecutive rounds without receiving a schedule, a node infers that it is no longer connected to the multi-hop network. It then attempts to join the single-hop VSN by listening during the single-hop schedule slot based on the information about the single-hop VSN included in the last multi-hop schedule it received. Upon successfully receiving the schedule, the node joins the single-hop VSN. Otherwise, the node moves to the bootstrapping VSN. Equivalently, the host optimizes the schedule of the multi-hop VSN by removing data slots associated with nodes from which it has not received any packets for $p$ rounds in a row. $p$ is a design parameter and can be chosen to reflect the duration for which a node remains synchronized to the global time without receiving any packets, the expected reliability with which energy harvesting nodes operate, and the strength of short-term fluctuations of the RF environment.

*Single-hop to Bootstrapping VSN.* Nodes in the single-hop VSN move to the bootstrapping VSN after $p$ consecutive rounds without receiving a schedule. The host also optimizes the schedule of the single-hop VSN by removing data slots assigned to nodes from which it has not received any data for $p$ rounds.

*Single-hop to Multi-hop VSN.* Lastly, nodes in the single-hop VSN transition to the multi-hop VSN if they are able to receive the schedule of the multi-hop VSN and are thus connected to the currently existing multi-hop network. Every $m$ rounds, the host indicates in the single-hop schedule that nodes should listen for the subsequent multi-hop schedule slot. The single-hop schedule also includes timing and networking information for the multi-hop VSN. Based on this information, all nodes in the single-hop VSN simultaneously listen for the multi-hop schedule every $m$ rounds. Any node that receives the multi-hop schedule immediately participates in the schedule's flood, allowing all nodes in the single-hop VSN that are able to join the multi-hop VSN to do so in the same round, see also the properties of E-WAN described in Section 4. The parameter $m$ presents a trade-off between how rapidly nodes move to the multi-hop VSN once they are connected to the multi-hop network and the overhead associated with scanning for multi-hop network traffic.

An example of E-WAN operating in a small energy harvesting network is shown in Figure 7. The energy harvesting nodes communicate in all three VSNs and transition between them to dynamically adapt to changing resources and network states.

## 4 PROPERTIES

In this section, we present properties of E-WAN related to energy efficiency and adaptability, which are essential for wireless communication in energy harvesting wide-area networks. The properties hold despite the proposed protocol's limited reliance on information about the network state and arbitrary spatial and temporal variability of energy availability at the nodes.

### 4.1 Energy efficiency

E-WAN enables energy harvesting nodes to efficiently communicate and effectively utilize their limited and time-varying resources. E-WAN achieves this through the following two properties.

The first property states that E-WAN allows nodes to locally optimize their resource requirements by dynamically adapting to changes in the environment and network state.

PROPERTY 4.1. *Assume that the energy requirement of long-range single-hop communication of a data packet is higher than that of short-range multi-hop communication for each size of single-hop and multi-hop VSN. Then with E-WAN, every node in the network locally optimizes its energy requirement to communicate data.*

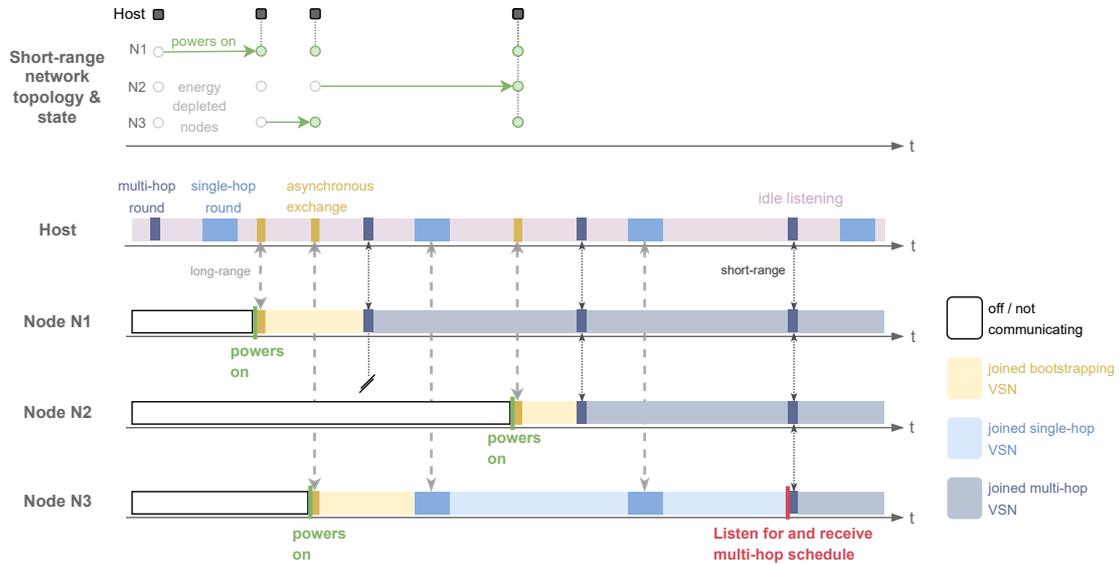

Fig. 7. The host is always part of all three VSNs and each energy harvesting node is in one of the four states depicted in Figure 4. Initially, the energy harvesting nodes are depleted. Once nodes accumulated enough energy to communicate, they join the bootstrapping VSN. With long-range communication, they asynchronously exchange a packet with the host to receive network and timing information of the subsequent multi-hop and single-hop communication, see Figure 6. This information enables them to scan for the schedule of first the multi-hop and then the single-hop VSN. Node N1 receives the multi-hop schedule and joins the multi-hop VSN. However, node N3 is unable to receive the multi-hop schedule because it is disconnected from the multi-hop network. Node N3, therefore, listens for the single-hop schedule slot and joins the single-hop VSN. When node N2 bootstraps, the network state enables it to join the efficient multi-hop VSN. Since node N3 is now connected to the multi-hop network, it receives the multi-hop schedule the subsequent time it listens for it and transitions to the multi-hop VSN.

The assumption states that the single-hop VSN's communication is always more energy-intensive than the multi-hop VSN's. Therefore, nodes locally optimize their communication energy by using the more efficient multi-hop VSN whenever possible. In E-WAN, nodes that are bootstrapping first attempt to join the multi-hop VSN, the more energy-efficient sub-network. Only when they are out of range of the multi-hop network do they join the single-hop VSN. In addition, nodes in the single-hop VSN periodically sample the state of the multi-hop VSN, and if they determine that they are connected to the multi-hop network, they move to the multi-hop VSN. By participating in the multi-hop VSN, they also extend the multi-hop network, potentially allowing other nodes to connect to it. There can be a delay of at most $m - 1$ rounds for a node to reach this local energy-efficient state after a multi-hop path to it has formed, since nodes sample the multi-hop communication periodically every $m$ rounds. Delays in forming a multi-hop network are discussed in more detail in Section 4.2. Conversely, nodes in the multi-hop VSN only move to the single-hop VSN if they lose connection to the multi-hop VSN's network. Furthermore, if nodes run out of energy to communicate in the multi-hop VSN, they also do not have the resources to do so in the single-hop VSN because of its higher energy demand. Nodes, therefore, rely on short-range multi-hop communication whenever possible and consequently locally optimize their resource requirements to communicate.

It should be noted that nodes joining the multi-hop VSN may increase the resource burden for other nodes in the multi-hop VSN. When the joining node has data to communicate and the host assigns data slots to it, other nodes in the multi-hop VSN idle listen during these data slots and, if they receive data packets, they retransmit them. This can adversely affect the operation of nodes in the multi-hop VSN by depleting their available energy. Nonetheless, when the energy requirement per relayed packet in the multi-hop VSN is relatively low, the energy benefit for the joining node to communicate in the multi-hop VSN rather than in the energy-intensive single-hop VSN, usually outweighs the negative impact on others. More importantly, dynamically changing energy and link budgets in conjunction with highly unreliable estimations of these quantities imply that a global network-wide optimization is not feasible.

As a consequence of Property 4.1, in a simple scenario where all nodes have sufficient energy and form a connected multi-hop network, energy harvesting nodes' communication converges to rely only on short-range multi-hop communication.

PROPERTY 4.2. *Suppose that the short-range link budgets are static and allow all nodes to communicate in the multi-hop VSN. Moreover, all nodes have sufficient energy to communicate at the intended packet rate. Then all energy harvesting nodes converge to the multi-hop VSN where they only rely on efficient short-range communication.*

There exist nodes that have a direct short-range link to the host, as otherwise, a multi-hop network that includes the host would not exist. According to Property 4.1, they join the multi-hop VSN. Furthermore, they remain in the multi-hop VSN as they have sufficient energy and the communication links are stable according to the assumptions of Property 4.2. Recursively, nodes that have short-range links to the currently existing network in the multi-hop VSN also join the multi-hop VSN according to Property 4.1. Similarly, these nodes remain in the multi-hop VSN according to the assumptions of Property 4.2. Therefore, all nodes eventually join and, due to the dependable short-range links and their sufficient resources, remain in the multi-hop VSN. Thereby, all energy harvesting nodes in the network converge to the multi-hop VSN forming a single multi-hop sub-network.

A consequence of the two properties above is that node connectivity does not need to be determined at design time or identified at runtime. Instead, E-WAN ensures that each node autonomously and independently chooses the VSN with the lowest energy demand for communication that permits the node to communicate.

## 4.2 Dynamic Behavior

E-WAN addresses fluctuations in the network state and the unreliable operation of energy harvesting nodes through dynamic transitions between VSNs. Nodes adapt to their local resources, and multi-hop topologies are exploited where and when possible, while nodes can rely on reliable communication otherwise. Nonetheless, the dynamic behavior incurs only a small overhead and does not require nodes to maintain any network state information. We summarize properties related to the adaptability that the protocol enables under the assumption that there are no packet losses when link budgets are sufficiently high for successful packet reception and that bootstrapping is successful every time.

When a node no longer has a multi-hop path to the host in the multi-hop VSN, meaning it is disconnected from the multi-hop network, it swiftly moves to the single-hop VSN, where it leverages reliable point-to-point communication with the host.

PROPERTY 4.3. *Suppose that a multi-hop round begins at time $t$, the round period is $T$, the time difference to the next single-hop round is $\Delta T$, and a node infers that it is no longer connected to the multi-hop VSN's network if it does not receive the schedule for $p$ consecutive rounds. Assume now that* *a node received the schedule of the multi-hop round starting at $t - T$, but after time $t$ there is no path to the host via the multi-hop VSN. Then, the node starts communicating via the single-hop VSN at time $t + p \cdot T + \Delta T$, if it has not run out of energy.*

Since the node received the schedule of the multi-hop VSN at time $t - T$, a multi-hop path to the node existed at that time. Subsequently, for the multi-hop rounds starting after time $t$, the node is unable to receive the schedule. $p$ communication rounds after $t$, the node has missed $p$ schedules and infers that it is no longer connected to the multi-hop VSN's network. It then scans for the schedule of the single-hop VSN's subsequent round at $t + p \cdot T + \Delta T$. Under the assumption that there are no packet losses, the node receives the single-hop schedule and switches to the single-hop VSN. The node is thus able to communicate again $p \cdot T + \Delta T$ after the multi-hop round at time $t$, for which it no longer had a multi-hop path.

In summary, the time interval during which a node cannot communicate with the host is bounded by the constants $p$ and $\Delta T$, and is independent of any other network state. A similar result can be obtained for the transition from the single-hop VSN to the multi-hop VSN.

PROPERTY 4.4. *Suppose at time $t$ a multi-hop round begins and the following holds: A non-empty subset $S$ of energy harvesting nodes is in the single-hop VSN. Nodes in the single-hop VSN sample the multi-hop VSN every $m$ rounds and the round period is $T$. Starting at time $t$, the nodes in the multi-hop VSN and a subset of the nodes $M \subset S$ could jointly form a connected multi-hop network, i.e., there is a multi-hop path from the host to each node in $M$. Then all nodes $M$ join the multi-hop VSN at the latest at time $t + (m - 1) \cdot T$, if they have not run out of energy.*

All nodes in the single-hop VSN sample the multi-hop VSN during the same round as instructed by the host. Any node in the single-hop VSN that receives the schedule of the multi-hop VSN round participates in the flood of that schedule, thus immediately extending the topology of the multi-hop VSN. Since nodes in $M$ and the nodes in the multi-hop VSN could form a connected multi-hop topology after $t$, all nodes in $M$ transition to the multi-hop VSN in the same round, namely when they sample the multi-hop VSN. Nodes in the single-hop VSN are instructed to sample the multi-hop VSN every $m$ rounds. However, this may occur in arbitrary relation to the round at time $t$. Therefore, in the worst case, nodes in the single-hop VSN sampled the multi-hop VSN state at time $t - T$ and thus only determine that there is a multi-hop path during the next sampling, $m$ rounds later. Consequently, all nodes in the subset $M \subset S$ in the single-hop VSN join the multi-hop VSN at most $(m - 1) \cdot T$ after the first multi-hop round at time $t$, when the larger multi-hop network in the multi-hop VSN can be formed.

In summary, the temporal behavior of dynamically extending multi-hop paths in the multi-hop VSN with nodes in the single-hop VSN is independent of the number of hops it is increased by and depends only on the constants $T$ and $m$. There is a trade-off between how often the multi-hop network traffic is sampled, which requires additional energy from nodes in the single-hop VSN, and how fast the multi-hop VSN's network builds up when possible.

The properties of E-WAN in terms of its dynamic behavior are important attributes for its adaptability and efficiency. Since the assumptions on packet losses and bootstrapping do not typically hold in a real-world deployment, energy harvesting nodes may, for example, need to attempt to bootstrap again at a later time by sending another synchronization request packet and thus spending a very small amount of additional energy. The nodes may therefore join the multi-hop VSN or single-hop VSN later. The consequences of packet losses depend on various communication parameters, such as the number of re-transmissions and the current network state. For instance, in the multi-hop VSN, the impact of a packet loss can vary depending on how many multi-hop paths there are to a node. Based on such factors, packet losses may have no or minimal effect on a node's operation, may lead to short-lived and superfluous transitions from one VSN to another, or may cause delayed transitions from the single-hop to multi-hop VSN. Nonetheless, we demonstrate the importance and benefits of E-WAN's dynamic behavior and energy efficiency in the detailed evaluation and real-world deployment presented in Section 6.

## 5 IMPLEMENTATION

We implement E-WAN on real hardware for a real-world case study, and we simulate the protocol for an extensive analysis in comparison to various baselines in different scenarios.

The implementation of the proposed protocol is based on the open-source Flora library [66] and is subsequently deployed on the DPP3e [48], a harvesting-based platform for advanced indoor sensing that has several distinct communication capabilities. The Flora library includes various radio drivers and protocol implementations and is compatible with the Semtech SX1262 transceiver. The DPP3e also integrates this transceiver [48].

The SX1262 supports both LoRa and FSK modulation schemes at a large range of transmission powers, thus offering link budgets of up to 170dBm [49, 64]. The numerous available configurations allow trade-offs between various aspects of communication, including link budgets, time-on-air (ToA), and achievable data throughput [49, 62]. Figure 8 provides insight into the wide range of communication capabilities of the SX1262 transceiver. It depicts the transceiver's sensitivity and ToA, determined with the Flora software [62, 64], for a selection of supported configurations. ToA is calculated for a packet with a payload of 20 bytes, a header, and cycling redundancy check (CRC). For LoRa, the bandwidth is 125 kHz, and for FSK, it is set according to the data rate (DR).

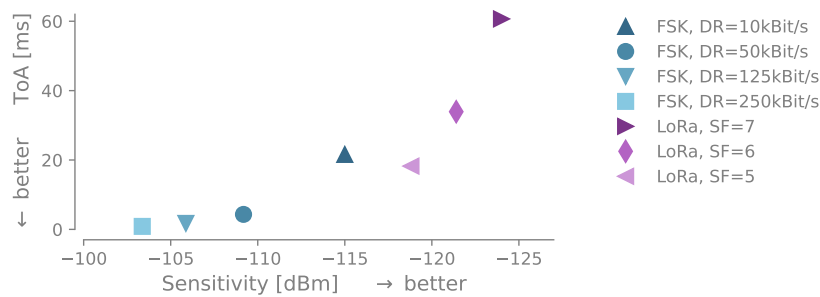

Fig. 8. The SX1262 has a large design space supporting both LoRa and FSK modulation schemes. Modulation scheme, data rate (DR), and for LoRa the spreading factor (SF) impact the time-on-air (ToA) and the receiver's sensitivity. The ToA is determined for a packet with a 20 byte payload, a header, and cycling redundancy check (CRC).

This transceiver is well-suited for the implementation of the proposed protocol. By exploiting different configurations, it allows for efficient short-range communication with small link budgets in the multi-hop VSN and large link budgets at a higher energy expenditure for communication in the single-hop VSN, supporting wide area networks. While various parameters could be explored, we focus on setting the modulation schemes and transmit powers, since they greatly impact the resulting communication range. With LoRa, the connectivity is on the order of multiple kilometers, and with FSK, it is on the order of covering a floor in a small building. Table 1 summarizes the configurations for the different VSNs and provides an overview of E-WAN's parameters.

Although all nodes have the same communication demand, this is not a requirement of the protocol. The long-range communication has a link budget that is 21 dBm larger than the short-range communication. This enables numerous application scenarios in which the energy harvesting nodes and the host form a network structure as shown in Figure 3. The overhead associated with listening for a schedule sent with FSK at $DR = 250\,kBit/s$ is orders of magnitude lower than the energy required for a communication round in the single-hop VSN, as reflected by the large difference in ToA in Figure 8. This dictates the trade-off between overhead associated with sampling the multi-hop VSN and how fast nodes transition to the multi-hop VSN when possible, see Section 4.2, and therefore, $m$ is set to a low value. The period of the communication is set to 5 min, which was identified in [57] as the longest period supported by similar hardware without nodes losing synchronization due to clock drift. Even with this long period, the energy available in indoor environments is typically insufficient to continuously support energy-efficient multi-hop communication [58].

Table 1. Communication parameter configuration for each VSN and an overview of the transition parameters and timing of E-WAN. Unless otherwise specified these parameters are used for all evaluations presented in Section 6. SF is the spreading factor, $F_C$ the center frequency, $BW$ the bandwidth, and $DR$ the datarate.

| VSNs & E-WAN Parameters | Parameter | Configuration |
|---|---|---|
| *Bootstrapping* VSN (Sync packet exchange) | Modulation scheme<br>Transmit power<br>Channel characteristics | LoRa with SF = 7<br>+ 14dBm<br>$F_C$=866.3125 MHz, $BW$=125 kHz |
| *Single-hop* VSN | Modulation scheme<br>Transmit power<br>Channel characteristics<br>No. of retransmissions<br>No. of hops<br>Traffic demand<br>Data payload | LoRa with SF = 7<br>+ 14dBm<br>$F_C$=863.3125 MHz, $BW$=125 kHz<br>0: energy harvesting nodes, 1: the host<br>1<br>All nodes request one slot per round<br>20 bytes |
| *Multi-hop* VSN | Modulation scheme<br>Transmit power<br>Channel characteristics<br>No. of retransmissions<br>No. of hops<br>Traffic demand<br>Data payload | FSK with $DR$ = 250 kBit/s<br>+ 14dBm<br>$F_C$=864.6875 MHz, $BW$=312 kHz<br>2<br>6<br>All nodes request one slot per round<br>20 bytes |
| Transitions | $p$<br>$m$ | 2 rounds<br>2 rounds |
| Timing | T<br>$\Delta T$ | 5 min<br>5 s |

The energy available at a node is directly tied to its environment and the efficiency with which a node harvests, stores, and supplies the energy to the application. The DPP3e platform relies on a boost and buck converter for efficient utilization of primary energy that has been converted to electrical energy [31]. The DPP3e features the AEM10941 harvesting chip with an ultra-low-power integrated boost converter and relies on the chip's maximum power point tracking (MPPT) to optimize the harvested energy. A TPS6274 buck converter is connected to the energy storage element and provides a stable voltage to the application circuit. The energy storage element is a CAP-XX supercapacitor with a usable capacity of B = 0.7 J. When the buck converter enables its output, the Apollo3 Blue Plus microcontroller turns on. Once a node determines that it has sufficient energy to start communicating, the communication subsystem, including the transceiver, is initialized, and the node enters the bootstrapping VSN. Numerous energy management methods have been proposed for adapting a node's behavior to its available resources [2]. We rely on a simple reactive approach and leave it to future work to optimize a node's energy management. A node determines that it has sufficient energy to start communicating when the usable energy in the energy storage surpasses the threshold of 0.115 J. This threshold is on the order as the resource requirements for bootstrapping and communicating for several rounds in the single-hop VSN. While waiting to accumulate enough energy to communicate, a node measures the supercapacitor's voltage $V_{\text{cap}}(t)$ every 30 seconds and estimates its stored energy according to $E_{\text{cap}}(t) = \frac{1}{2}CV_{\text{cap}}(t)^2$ where $C$ is the supercapacitor's capacitance. A node communicates as long as its buck converter supplies a stable output. While for our evaluation, all energy harvesting nodes are identical in terms of hardware and functionality, the protocol does not require such an assumption. Nodes could, for example, include different sensors. Such differences may impact how much of the harvested energy nodes have available for communication. Our experimental evaluation accounts for variable and diverse energy availability within the network.

### 5.1 Discrete event simulator

In addition to the implementation on real hardware, we implement the proposed protocol with the open-source discrete event simulator written by Roman Trüb *et al.* described in [64], and expand it to include an energy model of energy harvesting nodes.

The simulator's RF communication is modeled at a packet level. The modulation schemes, their link budgets, and the resulting ToA of packets are defined according to the Semtech SX1262 transceiver specifications [49]. A link between two nodes is defined by the link's path loss, and a network is specified by a path loss matrix. A node does not receive a packet if the signal power at the transceiver is lower than its sensitivity. For signals with sufficient power, the reception is probabilistic, with the probability of a failed reception increasing the closer the signal power is to the receiver's sensitivity. Packet reception is furthermore based on models and insights from works examining LoRa, concurrent transmissions, and the capture effect [14, 69]. The reception of an entire message, consisting of header and payload, is impacted by any transmission that temporally overlaps with the message's transmission. The probability of receiving one of two overlapping transmissions is specified as a continuous normal distribution and depends on the received signal powers of the two transmissions. The parameters of the simulated communication are set according to the overview in Table 1.

The simulator is extended to incorporate a simple, yet realistic energy model of an energy harvesting node [56]. The evolution of the energy storage state of charge is modeled according to

$$E_{\text{cap}}(t+1) = \max\left(\min\left(E_{\text{cap}}(t) + E_{\text{harv}}(t) - E_{\text{used}}(t), B\right), 0\right)$$

where $B$ is its maximum usable energy capacity, $E_{\text{harv}}(t)$ the energy the node harvests, and $E_{\text{used}}(t)$ the energy it uses from time $t$ to $t + 1$. The harvested energy is specified as a harvesting trace for each node, enabling simulations of different scenarios and environments. The energy consumption model takes into account a node's power requirement during radio operations, while it is sampling the supercapacitor voltage every 30 s, $P_{\text{boot}}$, in deep sleep mode between communication rounds $P_{\text{sleep}}$, and in idle mode between the slots of a round and, where applicable, retransmissions of packets, $P_{\text{idle}}$. Moreover, the energy demands when the application is turned on, $E_{\text{boot}}$, and when the communication is initialized, $E_{\text{com, init}}$ are considered. Resource requirements for other state transitions and for computing and processing are not integrated, as these take orders of magnitude less time than the considered tasks. The resources consumed by the radio during its operation, provided by the open-source simulator, correspond to the characterization for various transceiver configurations and modes of an equivalent platform. We characterize the DPP3e platform to determine the other parameters of a node's energy consumption model.

*Node Characterization.* We characterize the application circuits of two DPP3e nodes through a series of short experiments to estimate their energy consumption parameters. Current and voltage measurements are performed with the Rocketlogger [52]. Table 2 provides an overview of the measured power and energy consumption. Energy values are averaged across 50 samples and power numbers across measurements spanning a total of 20 min. The parameters are subsequently scaled to incorporate the buck converter's efficiency $\epsilon$, for which we use a value of $\epsilon = 0.9$, corresponding to a typical value for the typical operating range of energy harvesting nodes [33].

Table 2. Experimentally measured energy and power parameters of two DPP3es utilized in the energy model of the discrete event simulations.

| **Parameter** | $E_{\mathbf{boot}}$ | $E_{\mathbf{com, init}}$ | $P_{\mathbf{boot}}$ | $P_{\mathbf{sleep}}$ | $P_{\mathbf{idle}}$ |
|---|---|---|---|---|---|
| Experimentally measured value | 13.655 $\mu$J | 17.25 mJ | 27.254 $\mu$W | 26.831 $\mu$W | 10.516 mW |

## 6 EXPERIMENTAL EVALUATION

In this section, we demonstrate E-WAN's behavior and performance in terms of efficiency and dynamic adaptability compared to two baselines. A single-hop baseline relies solely on long-range point-to-point communication, and a multi-hop baseline where nodes exclusively employ efficient short-range multi-hop communication. The protocols are evaluated with event-based simulations for a number of different scenarios. Subsequently, we present a real-world case study in which we deploy a representative energy harvesting network and demonstrate the novel protocol operating in a real-world environment. The case study additionally verifies the validity of the simulations.

### 6.1 Discrete Event Simulations

With discrete event simulations, we extensively evaluate the proposed protocol's performance in comparison to two baselines. The simulations facilitate evaluating different energy harvesting scenarios and a wide range of network topologies. The different communication protocols are simulated under the same conditions to ensure comparable results.

*6.1.1 Baselines and Metrics.* We compare the proposed protocol to two baselines. The first baseline, referred to as the *single-hop* baseline, relies exclusively on point-to-point communication between each energy harvesting node and the host and is comparable to communication in LoRaWAN Class B networks [39]. Its communication is time-synchronized and follows periodic rounds. A round is equivalent to the single-hop VSN communication round structure described in Section 3.2. A node joins and starts communicating by sending an asynchronous packet to the host and receiving the timing of the subsequent communication round. This packet exchange also relies on point-to-point communication, though on a different channel. The parameters of the synchronization packet exchange and the communication in the rounds of this baseline are set identically to those of the bootstrapping and single-hop VSNs, as summarized in Table 1.

The second baseline, the *multi-hop* baseline, is the Low-Power Wireless Bus (LWB) protocol [27]. Its communication is equivalent to that of the multi-hop VSN, and nodes bootstrap by idle listening until receiving a multi-hop schedule. The communication parameters are set to the multi-hop VSN's parameters in Table 1.

The three protocols are compared with respect to several metrics. To this end, we define the total energy a node harvests throughout the simulation length $T_{\text{sim}}$ as $E_{\text{in}}$. Furthermore, the number of packets a node successfully sends throughout $T_{\text{sim}}$ is $P$. We define the total time during the simulation that a node has sufficient energy to communicate as $T_{\text{active}}$, spanning from when a node determines that it has sufficient energy to communicate until it runs out of energy. Lastly, we consider the communication rounds in which a node participates. These are the rounds for which a node successfully receives the first schedule sent by the host and participates in, irrespective of whether it successfully sends a packet. We denote the total time during the simulation spanned by such rounds until the node fails to participate in a round or runs out of energy as $T_{\text{com}}$. The following metrics are analyzed.

- *Efficiency:* The efficiency metric is the number of packets a node is able to successfully transmit relative to the energy it harvests. We define it as $P/E_{\text{in}}$. In the simulation setup, each node transmits packets with a 20 byte payload and has a traffic demand of one packet per round, see Table 1. This metric is affected by the resource requirements of communication, overheads incurred by bootstrapping, reliability, and a node's ability to communicate in different network states.
- *Liveness:* The liveness metric is defined as $T_{\text{com}}/T_{\text{sim}}$. It is the fraction of time that a node participates in the network's communication, i.e., receives a round's first schedule and participates accordingly, for either single-hop or multi-hop. In multi-hop rounds, the liveness of a node dictates the percentage of time the node is able to assist others in the network and relay their packets.
- *Downtime:* The downtime metric is defined as $(T_{\text{active}} - T_{\text{com}})/T_{\text{sim}}$. It is the fraction of time a node has sufficient energy to communicate but does not participate in a period's round. It summarizes a node's time spent bootstrapping and when missed schedules prevent it from participating in a round. It focuses on a node's downtime from the networking perspective. This metric is influenced by the ease with which nodes join a network, and communicate in different environments as well as under varying conditions.

*6.1.2 Scenarios.* We analyze the three protocols in four distinct scenarios, each with unique network topologies and characteristics. These scenarios capture a wide range of general topology-related challenges that communication protocols encounter, and their evaluation provides key insights into the protocols' behavior and performance. Moreover, these networking scenarios encompass numerous real-world applications of energy harvesting networks.

In each scenario, the network consists of fifteen energy harvesting nodes and the host. Furthermore, for all considered networks, every energy harvesting node has a direct point-to-point link to the host via long-range communication, i.e., LoRa modulation with *SF*=7. The network topology for short-range multi-hop communication, i.e., FSK modulation with *DR*=250 kBits/s, differs between the scenarios and is described in detail below. Unless otherwise specified, we use the open-source simulator's calculations of the path loss between the nodes based on their distance. As a result, the short-range topology is determined by the nodes' placement.

- Office Building (*OB*): The distribution of nodes aligns with the positions of nodes in Flocklab, a testbed for wireless communication in low-power networks [63]. The nodes are distributed throughout a floor of a small office building. The path loss matrix of the discrete event simulations is set according to the measured path loss between the nodes of the testbed. Most nodes have a direct short-range path to the host and many other nodes since the network only spans one floor of a small building. While link characteristics are based on real measurements from the testbed, and therefore reflect a concrete network realization, the evaluation of this scenario represents an example of a well-connected small network. Such network topologies arise when, for example, a limited indoor area is monitored with high spatial resolution, resulting in nodes being densely deployed within a limited area. For example, detailed air quality monitoring of an office floor in a building [15] can result in such a network.
- Bottleneck (*BN*): All short-range multi-hop traffic passes through a small number of nodes, specifically two, located close to the host. These two nodes, with their limited and variable energy availability, form a bottleneck for short-range multi-hop communication. Such a bottleneck can arise when the host is positioned at the edge of an otherwise well-connected network. Various industrial applications, such as monitoring smart manufacturing or logistics systems, may encounter a communication bottleneck if the host is located in a control room off to the side.
- Few Hops (*FH*): Similar to the OB scenario, all energy harvesting nodes are clustered closely around the host. The shortest short-range multi-hop path from any harvesting-based node to the host is at most two hops long. Such networks can arise in similar scenarios as the OB scenario, yet nodes are less densely deployed and cover a larger area.
- Many Hops (*MH*): In this scenario, the multi-hop path of several harvesting-based nodes to the host requires multiple hops. In the considered topology, all nodes are reachable from the host in at most five hops. For networks that expand more in one dimension than the other two, such a topology can result from placing the host at the end of the longer dimension. For example, in the detailed monitoring of a medieval tower in Italy across several floors [20], placing the host on the first floor could result in such a network topology. Another example is a monitoring network in an aircraft, as considered in [12], where the host is placed at the front or the end of the aircraft.

The short-range multi-hop topologies of the four scenarios are depicted in Figure 9.

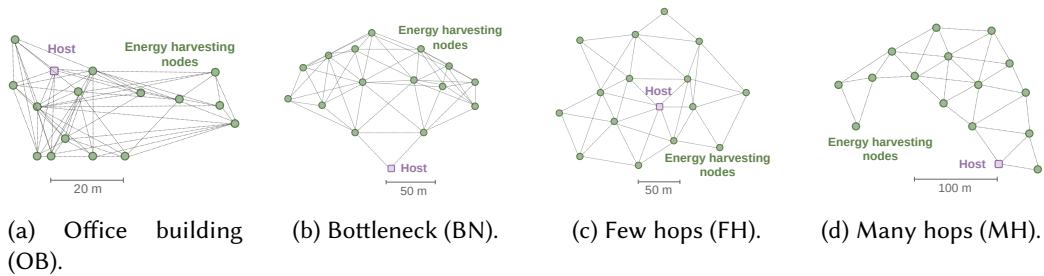

(a) Office building (OB).    (b) Bottleneck (BN).    (c) Few hops (FH).    (d) Many hops (MH).

Fig. 9. We analyze four scenarios with different short-range multi-hop network topologies. For all scenarios, long-range communication supports point-to-point communication from each energy harvesting node to the host.

6.1.3 *Energy Harvesting Environment.* We generate artificial energy harvesting traces for each node to evaluate the scenarios described in Section 6.1.2. For the real-world case study in Section 6.2, the corresponding discrete event simulation is based on harvesting traces measured at each node in the network.

The harvesting traces follow simplified day-night cycles. Nodes harvest energy for a limited number of hours during the day, and their harvested energy is zero otherwise, e.g., during the night. A node's average energy input per day, $E_{\text{avg}}$, is a random sample drawn from a uniform distribution ranging from 1 J to 10 J. The range represents various indoor environments in office spaces, from locations primarily exposed to artificial lighting to those receiving significant levels of natural sunlight [53]. The time window with input energy starts in the morning at a randomly sampled time, drawn uniformly between 05:00 and 10:00. Energy is harvested until the evening, with the end time drawn from a uniform distribution spanning from 16:00 to 21:00. During this time period, the harvested energy follows on average $E_{\text{avg}}$ and exhibits short-term hourly variability, modeled as a multiplicative noise term drawn from a normal distribution with zero mean and standard deviation $\sigma = 0.1$.

A network's performance is not only impacted by the available resources of individual nodes but, depending on the protocol, can also be influenced by how the harvesting characteristics at different nodes relate to one another. For example, it can matter whether nodes harvest the same amount of energy during similar hours or whether there is less temporal overlap between their energy inputs. We specify the network-wide characteristics of harvesting traces by drawing the random samples for the average daily energy $E_{\text{avg}}$, as well as the start and end time of input energy on a day of different nodes according to a specified correlation, $\rho$. We generate traces and analyze the scenarios in Section 6.1.2 for highly correlated harvesting traces, i.e., generated with correlation $\rho = 0.95$. Such correlated harvesting traces can result from nodes being located in the same large space or in similar proximity to windows with the same orientation. Furthermore, we investigate the performance of the protocols when the traces have more distinct characteristics, i.e., when they are constructed with a correlation of $\rho = 0$.

Figure 10 depicts samples of harvesting traces for each node in the network when the correlation is $\rho = 0$ and $\rho = 0.95$, respectively. The subsequent results are based on simulations over a seven-day time horizon. The results presented in the following sections are obtained from 20 Monte Carlo simulations, and the same harvesting traces are used when comparing two protocols.

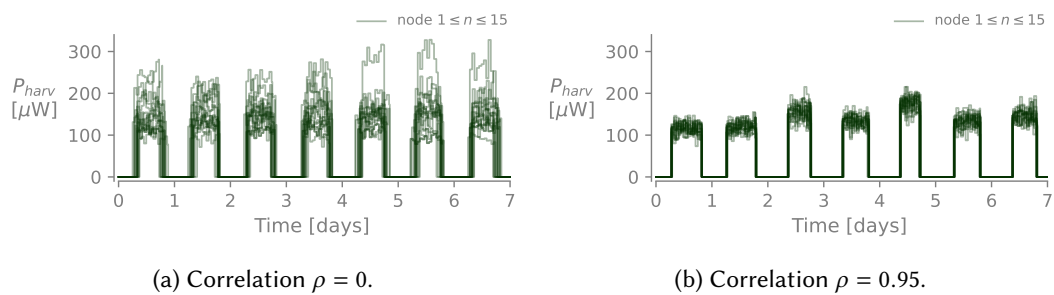

(a) Correlation $\rho = 0$.  (b) Correlation $\rho = 0.95$.

Fig. 10. The harvesting traces of nodes differ substantially when their parameters are random samples drawn with correlation $\rho = 0$. Conversely, when the correlation is high, i.e. $\rho = 0.95$, the harvesting traces of nodes only have small differences. The correlation of input energy between nodes influences their communication when they depend on one another.

6.1.4 *Energy-Efficient Bootstrapping.* We first evaluate E-WAN's bootstrapping mechanism which facilitates nodes in efficiently joining not only a single-hop but also a multi-hop network. To this end, we focus on the bootstrapping and multi-hop VSN, initially not considering the single-hop VSN, and compare this part of E-WAN to the *multi-hop* baseline. A more extensive evaluation of an equivalent bootstrapping mechanism was presented in [58]. The bootstrapping mechanism was compared to two baselines, a single-hop baseline with an asynchronous bootstrapping mechanism and a multi-hop baseline where nodes bootstrap by idle listening. The evaluation included both measurements and simulation results.

We denote the combination of only the bootstrapping and multi-hop VSNs as *DRB*, as in [58]. Since the single-hop VSN is not included, nodes in the multi-hop VSN transition to the bootstrapping VSN if they miss $p$ consecutive schedules. Furthermore, if nodes in the bootstrapping VSN do not receive the multi-hop VSN schedule, they do not attempt to join the single-hop VSN but instead bootstrap again at a later time with another synchronization packet exchange.

The energy storage capacity of the energy harvesting nodes is increased to 6 J, and for the multi-hop baseline, the threshold the reactive energy management employs to determine when to start communicating is set to 5 J. A higher threshold and energy storage capacity are necessary to enable nodes to idle listen for up to one period while bootstrapping.

The results of the many hops (MH) scenario for harvesting traces with $\rho = 0$ are shown in Figure 11. The distribution of each metric for each energy harvesting node over the 20 simulation runs is depicted.

The efficiency and liveness of DRB and the multi-hop baseline are similar for nodes one hop from the host, but for all other nodes, DRB significantly outperforms the multi-hop baseline. The performance of the multi-hop baseline degrades particularly drastically for nodes several hops away from the host. This effect is so pronounced that some nodes are entirely unable to be connected to the network and participate in communication rounds, i.e., their liveness is 0 %, and therefore so is their efficiency. Furthermore, for most nodes that are only a few hops from the host, the efficiency and liveness metrics exhibit greater variability, i.e., the interquartile range spans a wide range, for the multi-hop baseline compared to DRB. However, nodes several hops from the host have higher downtimes with DRB than with the multi-hop baseline.

In DRB, nodes with a direct short-range link to the host spend few resources bootstrapping since most of the bootstrapping time is spent in deep sleep, see also Section 3. Furthermore, the energy-efficient bootstrapping mechanism exhibits little variability in its resource demands [58]. These nodes are able to communicate in the multi-hop rounds regardless of the state of other energy harvesting nodes. This leads to high efficiency and liveness values and low downtimes. These nodes perform similarly with the multi-hop baseline. On the one hand, the threshold for starting to communicate is high to account for the high and variable bootstrapping energy cost when bootstrapping by idle listening [58]. But nodes often require less energy to join when they successfully bootstrap. Therefore, once joined, they have the resources to support communication for extended periods of time. On the other hand, if only a few nodes are communicating, as the liveness metric in Figure 11b demonstrates, the burden to forward messages from other nodes is small. These effects balance the improvements of DRB for nodes with a short-range link to the host, resulting in similar performances. Nonetheless, with the multi-hop baseline, the efficiency and liveness of these nodes exhibit larger variability because even small differences in the time it takes to bootstrap result in large differences in bootstrapping resource requirements due to the considerable average power consumption while idle listening for a schedule.

Due to the high energy cost of bootstrapping in the multi-hop baseline, the limited harvested energy only supports infrequent bootstrapping attempts, on the order of one per day. Therefore, nodes more than one hop from the host are only able to join the network and communicate if one of these seldom bootstrapping attempts occurs precisely when all nodes they depend on for forwarding are communicating. This becomes increasingly challenging to fulfill with increasing hop distance to the host, resulting in the drastic performance decreases in efficiency and liveness visible in Figure 11. When bootstrapping is unsuccessful, these nodes quickly run out of energy due to the significant resource requirements of idle listening. This leads to minuscule downtimes. With DRB, nodes that do not have a direct short-range link to the host, repeatedly exchange asynchronous long-range packets with the host to receive networking information and listen for the multi-hop schedule until they receive a schedule or run out of energy. Since they spend most of

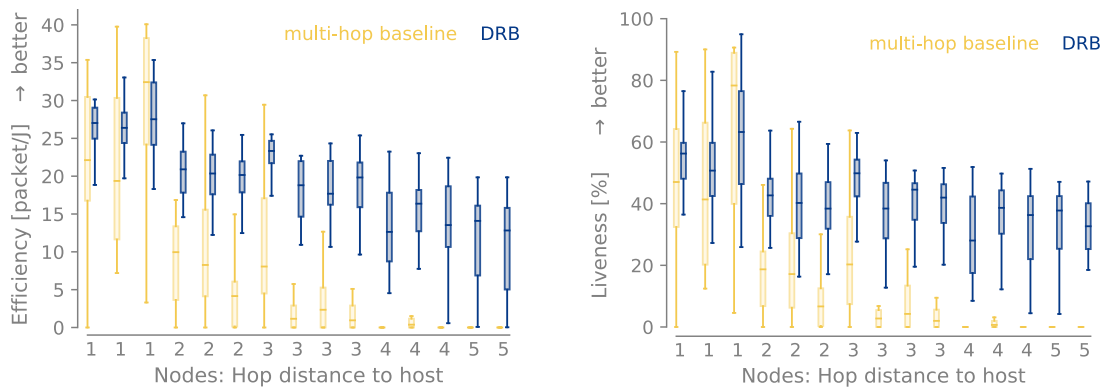

(a) The energy-efficient bootstrapping of DRB enables nodes several hops away to spend less energy on bootstrapping and send more packets.

(b) With DRB, nodes several hops away repeatedly try to join the multi-hop VSN. Once they are connected to the multi-hop network and join the multi-hop VSN, they still have the resources to participate in communication rounds.

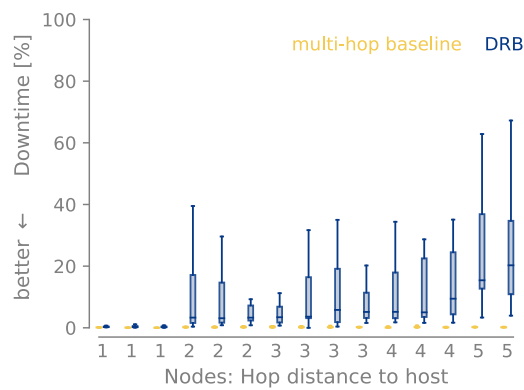

(c) Nodes in DRB are in the bootstrapping VSN and unable to communicate until they are connected to the multi-hop network or run out of energy.

Fig. 11. The distribution of the efficiency, liveness, and downtime metric for each node in the network over 20 simulations in the *Many Hops* scenario with harvesting condition $\rho = 0$. Combining E-WAN's bootstrapping and multi-hop VSNs (DRB) enables nodes to efficiently bootstrap. With this, nodes can attempt to bootstrap more frequently than with the multi-hop baseline and have more resources available to communicate.

the bootstrapping time in deep sleep, bootstrapping attempts are efficient, and nodes are able to try to join the multi-hop VSN frequently. Resources spent on more bootstrapping attempts lower the efficiency of nodes as their hop distance to the host increases. Furthermore, the liveness of nodes several hops from the host is bound by the liveness of nodes they depend on for connectivity to the multi-hop network. Nonetheless, with DRB, these nodes have significantly higher efficiency and liveness metrics than with the multi-hop baseline. Although the efficient bootstrapping VSN enables nodes that are at least two hops from the host to repeatedly probe and attempt to join the multi-hop VSN, these nodes are unable to communicate during this time even though they have the resources to do so. With only the bootstrapping and multi-hop VSNs, these nodes cannot adapt to various network states and effectively utilize their resources, leading to higher downtimes.

Table 3 summarizes the results of comparing DRB with the multi-hop baseline for all scenarios and both harvesting environments, i.e., $\rho = 0$ and $\rho = 0.95$.

| Metric | Protocol | Scenario, harvesting condition | | | | | | | |
| --- | --- | --- | --- | --- | --- | --- | --- | --- | --- |
| | | $\rho = 0$ | | | | $\rho = 0.95$ | | | |
| | | OB | BN | FH | MH | OB | BN | FH | MH |
| Efficiency [packet/J] | multi-hop baseline | 12.7 | 7.2 | 10.1 | 7.3 | 11.0 | 7.7 | 9.9 | 8.3 |
| | DRB | 22.2 | 19.7 | 21.0 | 19.0 | 20.5 | 19.3 | 20.2 | 19.2 |
| Liveness [%] | multi-hop baseline | 28.8 | 15.9 | 22.5 | 16.4 | 25.0 | 17.9 | 22.5 | 17.8 |
| | DRB | 45.5 | 41.3 | 43.3 | 40.7 | 44.3 | 43.5 | 44.0 | 43.3 |
| Downtime [%] | multi-hop baseline | 0.17 | 0.24 | 0.39 | 0.21 | 0.18 | 0.28 | 0.39 | 0.24 |
| | DRB | 1.4 | 6.6 | 4.7 | 10.7 | 0.86 | 2.38 | 1.8 | 3.4 |

Table 3. Mean of the efficiency, liveness, and downtime across all nodes in the network for 20 simulations for the networking and harvesting scenarios described in Section 6.1.2, respectively Section 6.1.3.

With DRB, the efficiency and liveness of nodes are significantly improved in all scenarios and harvesting conditions. However, nodes more often have the resources to communicate but are unable to, i.e., have higher downtimes. The design of the bootstrapping VSN enables nodes to use fewer resources for bootstrapping and therefore have more resources available to communicate and send packets. These improvements are essential for networking in low-power energy harvesting networks. Nevertheless, nodes can only communicate if they are connected to the multi-hop network of the multi-hop VSN. E-WAN's single-hop VSN is designed to address this limitation. When network sparsity leads to disconnections in the short-range topology, nodes can leverage the single-hop VSN's reliable communication to communicate when they have sufficient resources but are not connected to the multi-hop network.

*6.1.5 Efficient Multi-hop Communication.* We compare E-WAN to the single-hop baseline that only relies on point-to-point long-range communication. While both offer reliable but energy-intensive communication and an efficient bootstrapping mechanism, E-WAN enables nodes to dynamically adapt to changing network states and leverage short-range communication when possible.

Table 4 summarizes the mean efficiency, liveness, and downtime of nodes in each scenario with both harvesting characteristics, $\rho = 0$ and $\rho = 0.95$, over 20 simulations for the single-hop baseline and E-WAN. E-WAN consistently has higher efficiency and liveness than the single-hop baseline. The downtime of nodes is very small with both protocols.

Since the single-hop baseline relies only on point-to-point communication, nodes are able to bootstrap and communicate regardless of the state of other nodes. This leads to negligible downtimes that are primarily related to bootstrapping times and probabilistic packet losses. However, the long-range modulation scheme necessary to support point-to-point communication between each node and the host is energy-intensive. Consequently, nodes spend more of their limited resources on each transmitted packet than with short-range communication, which impacts their efficiency and liveness. E-WAN also enables nodes to bootstrap and communicate irrespective of the current network topology formed by the powered nodes. Nodes therefore exhibit low downtimes as well. However, the multi-hop VSN and the transitions between VSNs, see Section 3.3, enable nodes to communicate not only with energy-costly long-range modulation but also with efficient short-range communication when possible. This leads to minuscule improvements in downtime, as nodes run out of energy less frequently and thus need to bootstrap less often. Communication in the multi-hop VSN also improves efficiency and liveness due to its lower resource requirements. E-WAN, therefore, makes it possible for nodes to optimize and reduce the energy demand for communication in accordance with the network state while maintaining the ability to communicate whenever they have the resources to do so, regardless of the state of other nodes in the network.

|  |  | **Scenario, harvesting condition** | | | | | | | |
| :---: | :---: | :---: | :---: | :---: | :---: | :---: | :---: | :---: | :---: |
| **Metric** | **Protocol** | $\rho = 0$ | | | | $\rho = 0.95$ | | | |
|  |  | OB | BN | FH | MH | OB | BN | FH | MH |
| Efficiency [packet/J] | single-hop baseline | 10.3 | 10.2 | 10.2 | 10.1 | 9.5 | 9.4 | 9.5 | 9.4 |
|  | E-WAN | 20.4 | 19.3 | 20.0 | 18.9 | 19.5 | 18.6 | 19.4 | 18.8 |
| Liveness [%] | single-hop baseline | 28.9 | 30.0 | 29.6 | 29.3 | 31.2 | 33.3 | 32.2 | 32.0 |
|  | E-WAN | 40.9 | 40.5 | 40.7 | 41.3 | 41.9 | 42.1 | 41.9 | 42.4 |
| Downtime [%] | single-hop baseline | 2.07 | 1.56 | 1.60 | 2.10 | 1.97 | 1.29 | 1.41 | 1.95 |
|  | E-WAN | 0.83 | 0.95 | 1.24 | 1.45 | 0.64 | 0.66 | 0.96 | 0.87 |

Table 4. Average performance metrics for 20 simulations for the networking and harvesting scenarios described in Section 6.1.2 and Section 6.1.3. E-WAN improves over the single-hop baseline by enabling efficient multi-hop communication when possible.

*6.1.6 Reliable Single-hop Communication.* Lastly, we evaluate the importance of E-WAN's single-hop VSN by comparing it to DRB, the combination of only the bootstrapping and multi-hop VSN, as evaluated in Section 6.1.4.

E-WAN and DRB are simulated for the *Many Hops* scenario described in Section 6.1.2. The harvesting traces are set such that the nodes with a short-range link to the host harvest little energy compared to all other nodes in the network. Nodes more than one hop away from the host have an average daily input energy that is six times greater than that of the nodes with a direct short-range link to the host. Furthermore, these nodes start harvesting energy several hours earlier and continue harvesting for several hours longer than the nearby nodes. These harvesting conditions are chosen to illustrate the importance of E-WAN's single-hop VSN. E-WAN's behavior and performance generalize to other energy availability distributions among nodes in a network where nodes further away face limitations in their ability to rely on multi-hop communication due to differences in the amount of harvested energy or the temporal characteristics of the energy availability. However, the differences between E-WAN and DRB are typically less pronounced.

Figure 12 shows the mean efficiency, liveness, and downtime of each node for the simulation setup described above. Nodes one hop from the host have similar performance for E-WAN and DRB. However, E-WAN improves both efficiency and liveness while decreasing downtime in comparison to DRB for nodes that are at least two hops away from the host.

In both E-WAN and DRB nodes with a direct short-range link to the host behave the same. Via the bootstrapping VSN, they directly join the multi-hop VSN where they generally communicate using efficient short-range communication until they run out of energy. Only in the unlikely event that probabilistic packet losses cause such a node to miss two consecutive schedules do they leave the multi-hop VSN while still having energy. This leads to high efficiency and negligible downtime. However, since they harvest only little energy, they communicate only for a small percentage of the time, i.e., they exhibit low liveness.

For short-range multi-hop communication, nodes that are at least two hops away from the host depend on these nodes to relay packets. With DRB, these nodes are bound by the nodes with

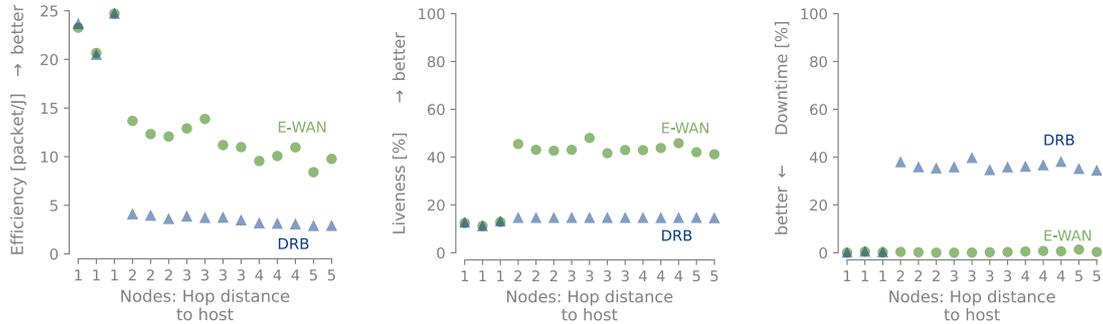

(a) E-WAN increases the efficiency of nodes several hops away as they communicate in the single-hop VSN when they are not connected to the multi-hop network instead of repeatedly bootstrapping.

(b) With E-WAN, the liveness of nodes several hops away is not bound by nearby nodes.

(c) The single-hop VSN enables nodes to communicate irrespective of the states of other nodes in the network.

Fig. 12. Results of simulating the *Many Hops* scenario when the nodes one hop from the host harvest significantly less energy than all other nodes. With E-WAN, nodes one hop from the host communicate only in the multi-hop VSN and thus have the same performance as when only using the bootstrapping and multi-hop VSN (DRB). E-WAN's reliable single-hop VSN enables nodes several hops away to communicate also when the nearby nodes are not powered improving their efficiency, liveness, and downtime.

a short-range link to the host since DRB only supports short-range multi-hop communication. Consequently, their liveness is dictated by that of the nodes one hop from the host and is also low. With the additional energy that the nodes further away from the host harvest, they are only able to repeatedly attempt to join the multi-hop VSN but not send more packets. They thus spend a significant percentage of time unable to communicate despite having the resources to do so. In contrast, with E-WAN, nodes several hops away can also join and communicate in the single-hop VSN when they are not connected to the multi-hop network. This enables them to send more packets and communicate whenever they have energy, significantly decreasing their downtime and improving their liveness. Since communication in the single-hop VSN is more energy-intensive than in the multi-hop VSN, their efficiency is not as high as that of the nodes one hop from the host, who communicate only in the multi-hop VSN. Nonetheless, with E-WAN, these nodes have an efficiency that is approximately twice as high as with DRB. The single-hop VSN of E-WAN is essential for enabling nodes further away to effectively utilize their resources and communicate even when nodes with a short-range link to the host are not powered.

## 6.2 Real-world Case Study

A small-scale energy harvesting network consisting of two energy harvesting DPP3e nodes and a host is deployed in a real-world indoor environment. This experiment demonstrates the protocol operating in a realistic setting. We furthermore experimentally measure the DPP3e nodes' state and energy-related evolution and compare the measurements to an equivalent simulation of the network under the same conditions. The high degree of similarity between the experimentally measured and the simulated network behavior shows that the simulations accurately represent an energy harvesting network and therefore validate the simulation results presented in Section 6.1.

*Setup.* The network consists of two energy harvesting DPP3e nodes [48] and a DPP2 LoRa platform [9] as a host. The nodes are placed in three rooms, forming a two-hop short-range network topology, as depicted in Figure 13. Both DPP3es have a direct link to the host with LoRa, $SF = 7$. Although small, the network and its topology enable measuring nodes in all three VSNs and observing transitions between them, described in Section 3.3. The communication parameters, including modulation scheme, transmit power, and transmitted data payload size, are set to the values described in Table 1. The communication period of the deployed network and the corresponding discrete event simulation is reduced to 3 min to account for the small network size.

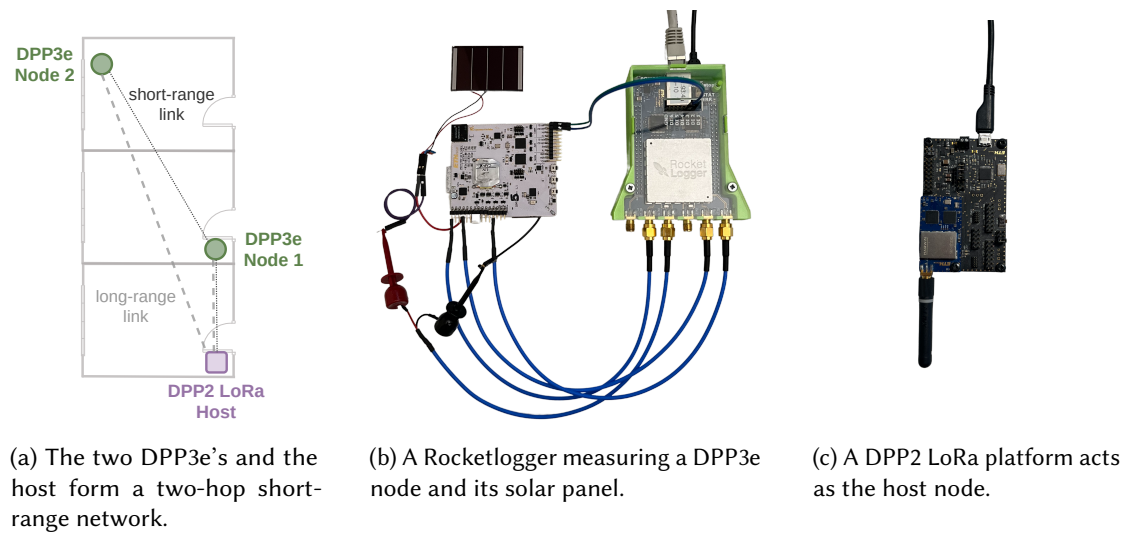

(a) The two DPP3e's and the host form a two-hop short-range network.

(b) A Rocketlogger measuring a DPP3e node and its solar panel.

(c) A DPP2 LoRa platform acts as the host node.

Fig. 13. The three nodes are set up to form a two-hop short-range network and each DPP3e has a direct long-range link to the host. An overview of the hardware shows the host, the DPP3e's, and the Rocketloggers that capture each energy harvesting node's harvesting power, supercapacitor voltage, and which VSN they are currently part of.

The behavior of the DPP3e nodes is measured with Rocketloggers [52]. We measure the current flowing out of the solar panel and the solar panel's voltage. The measured harvested power is used as the harvesting traces for the equivalent discrete event simulations. To account for inefficiencies related to the harvesting chip's maximum power point tracking (MPPT) and charging the supercapacitor, the harvesting traces are scaled by 0.85 in the corresponding simulation. A charging efficiency of 85 % lies within the range of typical efficiency values. The Rocketlogger furthermore records the supercapacitor's voltage and logs GPIOs that indicate the communication state of the node, i.e., whether it is communicating and which VSN it is currently in. The host logs the data packets it receives and their received signal strength indicators (RSSI). The RF environment is noisy, and in particular, the RSSI values of messages the host receives from node 2 in the single-hop VSN fluctuate considerably. The RSSI of packets the host receives from node 1 in the multi-hop VSN ranges from $-64$ dBm to $-67$ dBm, and the RSSI of messages from node 2 in the single-hop VSN varies between $-72$ dBm and $-85$ dBm. The network's behavior is measured over 12 hours.

*6.2.1 Results.* Figure 14 illustrates the results of the real-world case study. The experimentally measured harvested power of both DPP3e nodes is depicted. The harvesting traces employed in the simulation are a scaled version of the depicted input power. Furthermore, the supercapacitors' state of charge, derived from the experimentally measured supercapacitor voltage, in comparison to the simulated energy storage dynamics are shown. Lastly, each node's communication state, as indicated by the GPIOs, and the simulated state are visualized. Throughout the 12-hour experiment, the host logged the reception of 140 20 byte packets from node 1 and 117 from node 2. In the simulated network, the host received 134 20 byte packets from node 1 and 128 from node 2.

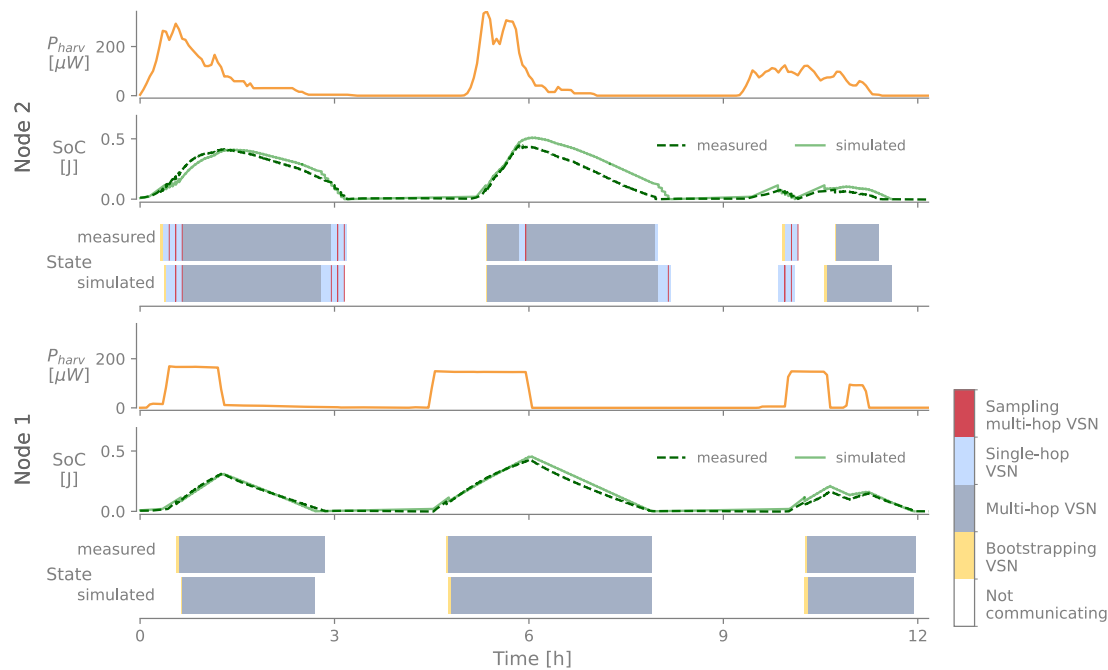

Fig. 14. The measured harvested power is scaled by 0.85 and provided to the discrete event simulation as harvesting traces. The resulting simulation and experimentally measured behavior of the DPP3e's closely align. The simulated and measured state of charge of each node are very similar. Furthermore, the order and timing of the communication including the transitions between VSNs align between the simulated and measured network.

*6.2.2 Analysis.* The two energy harvesting nodes harvest energy with distinct characteristics at different times. The harvested power of node 1 is typical for indoor environments dominated by artificial light, whereas the input power of node 2 follows smoother curves, which are more typical when a node is exposed to more natural light. In both cases, the input power is limited and exhibits variability that accurately represents indoor environments.

Node 2 harvests both more energy and earlier at the beginning of the experiment. When it bootstraps, it listens during the multi-hop VSN's schedule slot yet is unable to receive the schedule. The node subsequently listens for and receives the schedule of the single-hop VSN. It joins the single-hop VSN, where it communicates for several rounds and regularly samples the multi-hop VSN by listening for the multi-hop schedule slot. When node 1 has sufficient energy to communicate, it bootstraps and joins the multi-hop VSN due to its direct short-range link to the host. Thereafter, node 2 joins the multi-hop VSN as soon as it samples the state of the multi-hop VSN. When node 1 runs out of energy, node 2 misses 2 consecutive multi-hop schedules, listens for and receives the schedule of the next single-hop round, and thus transitions to the single-hop VSN, where it communicates until it runs out of energy. Hours later, node 1 accumulates enough energy to communicate again. It bootstraps and joins the multi-hop VSN. Subsequently, node 2 bootstraps and directly joins the multi-hop VSN since this time it was able to receive the multi-hop schedule when listening for it. Because of the noisy RF environment, the short-range path between node 1 and node 2 was temporarily unreliable around hour 6. This results in node 2 missing two schedules, moving to the single-hop VSN, and transitioning back to the multi-hop VSN when it receives the multi-hop schedule while sampling the multi-hop network state. Node 1 communicates in the multi-hop VSN until its energy depletes, at which point node 2 again transitions to the single-hop VSN, where it employs reliable long-range communication until it runs out of energy. Node 2 accumulates enough energy to begin communicating again before node 1. Due to its limited harvested power and the considerable resource demand in the single-hop VSN, it runs out of energy before node 1 rejoins the multi-hop VSN. When it bootstraps again, node 1 is part of the multi-hop VSN, enabling node 2 to directly join the multi-hop VSN, where both nodes communicate until they run out of energy.

The case study demonstrates E-WAN operating in a real-world environment and highlights its advantages. The energy harvesting nodes optimize their local resource requirements for communication by dynamically moving between the single-hop and multi-hop VSN based on realistically obtainable information. The nodes efficiently bootstrap, employ energy-efficient multi-hop communication when possible, and use reliable but energy-intensive single-hop communication otherwise. As such, the protocol enables nodes to adapt to varying resources and network state changes, thus efficiently utilize their resources.

Lastly, we compare the experimental measurements to the equivalent discrete event simulations. The simulated supercapacitor states of charge closely follow the measured ones. The order, transitions, and timing of the simulated communication states are also very similar to the measured states. Consequently, the simulated number of received packets differs only slightly from the number of packets received by the host during the experiment. The high degree of similarity between the experimentally measured and the simulated network behavior highlights that the simulations accurately represent the behavior of an energy harvesting low-power network.

At around 6 hours, the measured state of charge of node 2 experiences a decrease that is not reflected in the simulations. This occurs when node 2 temporarily moves to the single-hop VSN due to variations in the RF environment, increasing its resource requirements for communication. Although packet receptions are probabilistic in the simulations, node 2 did not exhibit the same short-term behavior, and the simulated energy consumption during those rounds is lower since the multi-hop communication is more efficient. This results in a minor shift in its simulated state of charge compared to the measured one until the node runs out of energy. Nonetheless, the comparison underlines the validity of the discrete event simulations and results presented in Section 6.1.

## 6.3 Limitations

E-WAN is a reliable and efficient networking protocol for energy harvesting networks. With the properties in Section 4 and the extensive evaluation in Section 6, we demonstrate key aspects of E-WAN. Nonetheless, it has some limitations. For the effective use of the bootstrapping and single-hop VSN, each node in the network needs a point-to-point link to the host. This limits the applicability of E-WAN to scenarios as depicted in Figure 3, where, with long-range communication, each node has a direct link to the host. Yet, with recent advances like LoRa, which has communication ranges of several kilometers in urban environments, this is feasible for a wide range of application scenarios.

Furthermore, the protocol enables diverse applications by supporting nodes with varying communication demands. However, the number of data slots that can be included in each multi-hop, respectively single-hop, communication round is bounded primarily by the communication period. The implementation in Section 5 is based on all nodes having the same communication demand, which can be met by one data slot per round. For more complex scenarios where nodes have widely varying communication requirements, over-provisioning available data slots in each VSN would be necessary. Otherwise, nodes switching between the single-hop and multi-hop VSNs may temporarily find themselves in a VSN that is unable to support their communication demand.

With E-WAN, nodes locally optimize their communication resource requirements by moving to the multi-hop VSN whenever possible, see Property 4.1. While this is important to address each node's limited and varying resources, it may result in nodes frequently transitioning between the single-hop and multi-hop VSN. The speed at which nodes can transition is bounded by the parameters $p$ and $m$, as outlined in Section 4. Small parameter values enable swift dynamic adaptation to changes in the network but may also lead to frequent transitions. Scenarios in which such frequent transitions occur include cases where the multi-hop path between the host and a node appears and disappears, e.g., due to fluctuating RF environments affecting nodes that are at the edge of the reach of the multi-hop sub-network. In addition to minor energy overheads associated with listening for missed schedules, frequent VSN transitions most notably impact the rounds in which the repeatedly transitioning node can communicate its data. This results from the $p$ rounds the node does not participate in while it misses the schedule. In addition, a node that joins a VSN first needs to inform the host of its traffic demand and can communicate only in the subsequent round if the communicated demand is incorporated into the schedule.

Furthermore, with the local optimization of communication resources, nodes joining the multi-hop VSN increase the communication energy burden of other nodes in the multi-hop VSN for their own reduced energy demand. E-WAN does not consider the complex network-wide state where, on the one hand, nodes in the multi-hop VSN may run out of energy if more nodes join the multi-hop VSN and, on the other, nodes in the single-hop VSN that have a lot of resources may be able to support long-range communication without risking energy scarcity in the future. Moreover, other nodes that have sufficient energy to communicate may or may not require these energy-rich nodes in order to be connected to the multi-hop network.

However, dynamically changing energy and link budgets, in conjunction with highly unreliable estimations of these quantities, imply that a global network-wide optimization is not feasible. As such, E-WAN focuses on each node managing its own resources and locally optimizing its energy demand for communication. We leave it to future work to expand on the presented idea and explore other aspects of the design of VSNs and their transitions.

## 7 RELATED WORK

In this section, we provide a summary of communication in LPWANs and a more detailed overview of communication for energy harvesting networks. Low-power wireless communication is a broad field that has been extensively investigated. Existing wireless communication technologies span a wide design space characterized by diverse link budgets and data rates [62]. For example, Bluetooth Low-Energy (BLE) or IEEE 802.15.4 cover short distances and provide limited data rates. Larger link budgets for low-bandwidth long-range communication can be achieved with LoRa or SigFox [46].

*Physical Layer.* Various protocols and algorithms have been proposed to adapt the transmit power in LPWANs that rely on short-range communication, examples include [34, 38]. For energy harvesting networks, transmit powers can be set according to communication characteristics and the nodes' resources, e.g., [5, 18, 36]. In [57], we also propose using increased transmit power to overcome network sparsity due to varying energy availability. Furthermore, related work has explored adaptive source coding, adaptive channel coding, and modulation scheme selection in energy harvesting networks [7, 18, 19, 40, 42]. Yet, determining appropriate parameters, such as the modulation scheme, can be difficult, in particular, without accurate energy predictions [2]. Consequently, several proposed algorithms assume perfect knowledge of the future harvested energy [2].

In contrast, E-WAN enables nodes to autonomously and locally optimize their resource requirements for communication. This optimization occurs by switching between long-range and short-range communication, e.g., modulation schemes and transmit power, based solely on local information.

*Protocols for Star and Star-of-Stars Topologies.* LoRaWAN is a common protocol for LPWANs with star-based topologies [39]. It offers variations for both asynchronous communication [1] and time-synchronized communication [39]. Various works build on LoRaWAN and optimize its performance for improved throughput, efficiency, reliability, or scalability, see for example [23, 37, 47, 65].

Energy harvesting networks can also form such topologies, facilitating the design and operation of networking protocols. In [8] and [13], energy harvesting nodes communicate with LoRa in star-based topologies over large distances. However, communication significantly contributes to the nodes' energy consumption. Conversely, in [54], harvesting-based nodes communicate directly with a smartphone with BLE, optimizing the utility of the transmitted data. The necessary direct link to the smartphone and use of BLE limit the distance over which nodes can successfully transmit their data. Furthermore, various MAC protocols for short-range energy harvesting networks with star topologies have been proposed [26, 41].

E-WAN also leverages point-to-point communication between each energy harvesting node and the host for reliable communication regardless of the state of other nodes in the network. This direct communication employs long-range communication and can thus cover wide areas. Yet, with the multi-hop VSN, it also enables nodes to rely on more energy-efficient communication when possible.

*Bootstrapping.* Different methods have been proposed to enable nodes, both energy harvesting as well as battery-based, to bootstrap. Nodes can incorporate additional hardware, such as wake-up receivers (WuRs), with which they continuously monitor network traffic. In multi-hop networks, nodes retransmit the wake-up signal to trigger nodes within their reach [61]. Although WuRs are designed for ultra-low power consumption [32, 44], they are prone to false wake-up, e.g., with a rate of $10^3/s$ [32]. Nodes may also rely on other signals to wake up, e.g., visible light [17], correlated measurements of physical events [11], and flickering in mains-powered artificial lighting [30]. Yet these approaches have limited applicability or scalability. Nodes in a network can also idle listen with their main transceiver until receiving relevant network traffic [24, 27, 66]. However, this method typically incurs a variable and substantial energy cost. In LoRaWAN Class B, nodes have a direct link to the gateway and they can receive information about the communication timing with a direct asynchronous packet exchange with the gateway [39]. This enables them to join a network efficiently since nodes can spend most of their bootstrapping time in a deep sleep state.

In [58], we propose an energy-efficient bootstrapping mechanism that relies on an asynchronous long-range packet exchange and enables nodes to efficiently join a short-range multi-hop network. This work builds on our previous energy-efficient bootstrapping approach for multi-hop networks and proposes a networking protocol for efficient and reliable communication in energy harvesting networks.

*Routing Protocols.* Numerous routing protocols for multi-hop networks have been proposed, an overview is presented in [3]. Since distributed embedded systems are generally resource-constrained, many protocols optimize resource requirements, balance load distribution, or improve network lifetime, e.g., [50, 51]. Typically, routing protocols require distributed network state information such as routing trees, link budgets, neighbor nodes, or network topology to facilitate routing decisions [3]. Numerous works have been dedicated to estimating the links in a network to provide protocols with the information they need [6].

Routing in energy harvesting networks has to take into account each node's limited resources, e.g., its varying energy availability. Energy harvesting is incorporated by considering factors such as the wasted energy due to each node's finite energy storage element, each node's predicted energy availability, residual energy availability, or a node's harvesting rate. A recent overview of routing strategies in energy harvesting networks is given in [2]. However, the unreliable operation of energy harvesting nodes due to the limited and unpredictable indoor photovoltaic energy availability, on the one hand, and variable RF propagation and typically slow convergence to collect all the necessary connectivity information, on the other hand, complicate the efficient operation and performance of routing-based protocols for energy harvesting networks in realistic indoor environments.

*Flooding-based Protocols.* Another class of protocols for multi-hop networks relies on synchronous transmissions, an overview of which is given in [71]. Synchronous transmissions can be exploited to flood the network with a packet, e.g., as in Glossy [28], RedFixHop [25], or BlueFlood [4]. Protocols like LWB [27] or Concurrent Transmissions with Forwarder Selector (CXFS) [16] build on these floods, whereby all nodes in a network or a subset of the nodes participate in a flood. Flooding-based protocols are characterized by low latency, high reliability, and high resilience.

The challenge in adopting flooding-based communication in energy harvesting networks lies in the importance of many nodes transmitting packets simultaneously, which is at odds with the temporal and spatial variability of energy availability. In [57], we nonetheless proposed a flooding-based protocol that builds on LWB for energy harvesting networks. To account for nodes having different energy availabilities, nodes with more energy communicate more frequently. Network sparsity is overcome by increasing the transmission power, but all communication relies on multi-hop communication. As such, [57] does not address situations where nodes are not connected to the multi-hop network. Furthermore, the bootstrapping mechanism relies on idle listening which is inefficient and detrimental in networks with several hops.

Conversely, in E-WAN, nodes bootstrap efficiently and rely on flooding-based short-range communication when possible. The availability of multi-hop communication is dictated by the network topology formed by the currently powered nodes and the host. Nonetheless, E-WAN does not require this topology to be determined at runtime. Nodes for which multi-hop communication is not possible employ reliable, albeit energy-intensive, communication. Reliable point-to-point communication with the host is achieved with long-range communication, e.g., with a long-range modulation scheme such as LoRa. In comparison to adapting only the transmit power as in [57], this approach offers significantly larger link budgets.

*Mixed-Range Protocols.* In [64], a protocol was proposed for networks with heterogeneous links. The protocol leverages a short-range and a long-range modulation schemes during a single periodic communication round. [29] designs a network consisting of nodes that only communicate with short-range modulation and nodes that contain hardware with both short-range and long-range capabilities. The former are grouped into clusters, while the latter act as cluster heads connecting the clusters to each other and to a server.

E-WAN also builds on combining short-range and long-range communication, achieved, for example, by using a single transceiver supporting different modulation schemes. It addresses the challenges of low-power networking in energy harvesting networks by exploiting long- and short-range communication for their respective properties. As such, E-WAN opens new possibilities for energy harvesting-based networks. Moreover, the general concept of E-WAN, which encompasses several virtual sub-networks with distinct trade-offs, effectively addresses challenges brought by network state changes, even when only limited information is available. This concept, therefore, has the potential to address communication challenges in a broader range of application scenarios beyond energy harvesting networks, including battery-based networks with mobile nodes.

## 8 CONCLUSION

We present E-WAN, a novel networking protocol for energy harvesting networks that is based on the concept of virtual sub-networks (VSNs). E-WAN enables nodes to efficiently use their available resources to communicate despite temporal and spatial energy variability, and thus, a changing network state. Nodes in the multi-hop VSN communicate with energy-efficient short-range multi-hop communication. When nodes are not connected to the multi-hop network, they can communicate in the single-hop VSN, which relies on energy-intensive long-range point-to-point communication. Another VSN is designed for bootstrapping, enabling nodes to efficiently join one of the other two sub-networks. Nodes transition between VSNs based on their local energy availability and receiving and missing schedules. The structure of the VSNs and their transitions enable us to provide crucial properties of the protocol in terms of efficiency and dynamic adaptability despite E-WAN's limited reliance on network state information. We furthermore implement the proposed protocol on real hardware and simulate the protocol for an extensive evaluation. We simulate various networking and harvesting scenarios, demonstrating the importance of the efficient bootstrapping mechanism, the efficiency of the multi-hop VSN, and the advantages the reliable single-hop communication provides when nodes are not connected to the multi-hop network. Furthermore, we experimentally illustrate the protocol operating in a deployment in a real-world setting. Under realistic harvesting conditions in indoor environments, nodes in a network fully leverage the characteristics and advantages of each VSN, as well as the transitions between VSNs, to efficiently communicate and effectively exploit their resources.

## ACKNOWLEDGMENTS

The authors wish to thank Roman Trüb and Reto Da Forno for their insightful discussions. This work was supported by the Swiss National Science Foundation under the NCCR Automation project, grant agreement 51NF40_180545.